\documentclass[11pt]{article}
\usepackage{pstricks}
\usepackage{color}
\usepackage[english]{babel}
\usepackage[spanish]{layout}
\usepackage[utf8]{inputenc}
\usepackage{layouts}
\usepackage{amsmath}
\usepackage{enumerate}
\usepackage{textcomp}
\usepackage{url}
\usepackage{epstopdf}
\usepackage{graphicx} 
\usepackage{subfigure}
\usepackage{cancel}
 
\newcommand{\bc}{\begin{center}}
\newcommand{\ec}{\end{center}}

\newcommand{\be}{\begin{eqnarray}}
\newcommand{\ee}{\end{eqnarray}}
\newcommand{\bi}{\begin{itemize}}
\newcommand{\ei}{\end{itemize}}
 
\author{\small E. A. Reyes R. $^{1}$ ~ A. R. Fazio $^{2}$\\
\footnotesize $^{1,2}$ \textsl{ Departamento de F\'{i}sica, Universidad Nacional de Colombia,}\\ \footnotesize \textsl{Ciudad Universitaria, Bogot\'{a} D.C., Colombia.}}
\date{\small \today}
\title{\textcolor{blue}{The Lightest Higgs Boson Mass of the MSSM at \\ Three-Loop Accuracy}}
\begin{document}
\maketitle

\begin{abstract}
In this article we provide explicit computations of the quantum corrections to the lightest  CP-even Higgs Boson mass in the real version of the Minimal Supersymmetric Standard Model. We compute the contributions coming from the SUSY-QCD sector with a precision of three loops at order $O(\alpha_t\alpha_s^2)$. The renormalization procedure adopted is based on the dimensional reduction scheme in order to preserve supersymmetry to all perturbative orders. The calculation extends the region of validity of previous studies to the whole supersymmetric parameter space. The involved master integrals have been computed by exploiting new proposed approaches based on the dispersion relation techniques for the numerical calculation of three-loop vacuum integrals with arbitrary mass scales.
\end{abstract}   

\section{Introduction}
\label{intro}
Since the discovery of the Higgs boson particle at the LHC and until the date there are no conclusive observations of new physics beyond the Standard Model (BSM). It seems mandatory the development of new initiatives focused on the search of data that prove the existence of new physical degrees of freedom. This is the main aim of projects at future colliders, like the International Linear Collider (ILC) \cite{ILC} and Future Circular Collider (FCC) \cite{FCC}. Even at $250~\rm{GeV}$ with a total integrated luminosity of $2~ab^{-1}$ there are proposals for the $e^+~e^-$ ILC collider to precisely measure the couplings of the $125~\rm{GeV}$ Higgs boson to vector bosons, quarks and leptons~\cite{Phys250} with an accuracy of $O(1\%)$ compared the $0.2\%$ accuracy based on the SM predicted couplings in terms of the Higgs boson mass. That augmented precision would allow to detect the small deviations of the typical BSM scenarios. In order for the theory to meet experiments a control of perturbation theory beyond two-loops level is required. The possible methods and strategies to match the experimental accuracy expected at the FCC-ee were recently discussed in~\cite{Blondel}. For instance the FCC-ee Tera Z for Electroweak Pseudo Observables would require three-loop electroweak calculations complemented with QCD four-loop terms \cite{TeraZ}. Quantum corrections must reach this level of accuracy in particular for the Higgs boson mass $M_h$ because it affects in a dominant way the value of all the observables involving the interactions of the Higgs. A very precise computation of $M_h$ including additional interactions coming from BSM scenarios could give us valuable hints about the existence of new physics. There are good reasons to prefer the Minimal Supersymmetric Standard Model (MSSM) as the BSM scenario which can describe these new interactions. According with the currently accepted analysis of the Higgs boson properties in the SM~\cite{Kniehl, Isidori, Edilson1, Edilson2}, the assumption that no new physics appears up to the Planck scale ($\Lambda_P$) renders the Higgs effective potential meta-stable and forces to accept an unnatural high amount of fine-tuning ($10^{34}$) for the squared Higgs boson mass at the electro weak scale ($\Lambda_{EW}$) leaving the hierarchy problem without solution. The MSSM can provide for a solution to this unusual fine-tuning by the existence of additional bosonic contributions that cancel the unwanted quadratic divergences. Moreover similar contributions render the effective potential stable, thus the MSSM also cures the vacuum stability problem. It is worth to mention that MSSM also provides a dark matter candidate, a mechanism to explain the neutrino oscillations and a platform to include gravitational interactions. \\

In most of the benchmark scenarios for MSSM Higgs boson searches, the Higgs boson found at LHC corresponds to the lightest CP-even Higgs boson with a mass $M_h$ which is not a free input parameter but it is predicted in the MSSM. The upper bound on its predicted mass at leading order (LO) is given by the Z gauge boson mass, $M_Z=91.2$~GeV, leading to the exclusion of the MSSM at current collider experiments. However, higher order quantum corrections to $M_h$ lead to a considerably large shift on its central value, $\Delta M_{h}\approx 40$~GeV, making the MSSM Higgs sector compatible with the mass and the detected production rates of the LHC Higgs-like signal over a wide range of the parameter space of the relevant phenomenology scenarios \cite{Carena2}. The state of art of the corrections to the Higgs boson mass in the MSSM is quite advanced and widely studied. At one-loop level the full corrections are found in~\cite{Dabelstein} with real parameters. The detailed results of a Feynman diagrammatic (FD) calculation of the leading two-loop QCD corrections at order O($\alpha\alpha_s$) can be found in~\cite{alphaalphas}, in particular the O($\alpha_b\alpha_s$)~\cite{alphabalphas} and O($\alpha_t\alpha_s$)~\cite{alphatalphas} contributions using the FD approach are known in the limit where the external momentum vanishes and in the MSSM version with complex parameters. In this limit there is an alternative procedure to compute the above corrections, the Effective Potential (EP) approach. A comparison of the corresponding two-loop results in the FD and EP approaches at O($\alpha\alpha_s$) can be found in \cite{Carena1, CarenaHollik} and references therein. In contrast to the EP method, the FD approach has the advantage that it can allow for non-vanishing external momentum. An evaluation of the momentum dependence of the two-loop corrections, including all the terms involving the QCD couplings, in the modified dimensional reduction scheme ($\overline{DR}$) was presented in \cite{MartinHiggs}. The latest status of the momentum-dependent two-loop corrections was discussed recently in \cite{Sophia, DegrassiMSSM} using a hybrid on-shell-$\overline{DR}$ scheme and including corrections of $O(p^2\alpha_t\alpha_s)$ for the real version of the MSSM. A complete two-loop QCD contributions to $M_h$ in the MSSM with complex parameters including the full dependence on the external momentum can be found in \cite{SophiaPaser}. Finally, at three-loop level there is a first diagrammatic computation performed by P. Kant and collaborators \cite{Harlander1, Harlander2, Kant}, where the radiative corrections to $M_h$ were computed in the SUSY-QCD sector including non-logarithmic terms of order $O(M_t^2\alpha_t\alpha_s^2)$. They have exploited the methods of asymptotic expansion in order to provide precise approximations in the relevant mass hierarchies. Their results were written in the publicly available code H3m and then implemented into the C++ module Himalaya \cite{Himalaya, Himalaya2} linked to the Mathematica generator FlexibleSUSY~\cite{FlexibleSUSY, FlexibleSUSY2} in a pure $\overline{DR}$ context. \\ 

With our work we provide an alternative calculation of the three-loop corrections to the lightest Higgs boson mass in the SUSY-QCD sector of the real version of the MSSM (rMSSM) at order O($\alpha_t\alpha_s^2$) \cite{Edilson3}. We have followed the Feynman diagrammatic (FD) procedure to obtain a renormalized correction in the $\overline{DR}$ scheme. Taking in mind that in the MSSM is no clear a priori what are the hierarchies of the masses, we have avoided the application of asymptotic expansions at the integral level and we have obtained a correction in terms of a set of three-loop master integrals whose numerical evaluation is possible for an arbitrary mass hierarchy thanks to the development of new calculation techniques based on the dispersion method \cite{Freitas1}. Our results are presented in this paper as follows. Section \ref{sec-1} contains a description of the renormalization procedure to derive the quantum corrections to $M_h$ coming from the Higgs sector of the rMSSM. The technical details of our computations are discussed in Section \ref{sec-2}. In Section \ref{sec-3} we present a numerical analysis where the effects of the three-loop corrections on the pole mass $M_h$ are evaluated at some kinematic limits. We give our conclusions in Section \ref{sec-4}.   

\section{\large Renormalization of the CP-Even Higgs Boson Masses}
\label{sec-1}

The Higgs sector of the real MSSM Lagrangian requires the definition of two doublets with opposite hyper-charges 
\begin{eqnarray}
H_{1}=\left(\begin{array}{c}
H_{1}^{0}+\frac{v_{1}}{\sqrt{2}}\\
H_{1}^{-}
\end{array}\right) & \rm{and} & H_{2}=\left(\begin{array}{c}
H_{2}^{+}\\
H_{2}^{0}+\frac{v_{2}}{\sqrt{2}}
\end{array}\right), \label{eq:Doublets}
\end{eqnarray}
with complex components that are vevless scalar fields, $H_j^0 = \phi_j^0 + i\chi_j^0$, from the expansion around the vacuum expectation values $v_{1,2}$. In terms of  $H_{1,2}$ doublets the expression of the Higgs sector of the bare Lagrangian is:
\begin{eqnarray}
\mathcal{L_{H}}=\partial_{\sigma}H_{1}^{\dagger}\partial^{\sigma}H_{1}+\partial_{\sigma}H_{2}^{\dagger}\partial^{\sigma}H_{2}-V(H_{1},H_{2}).\label{eq:LagHiggs}
\end{eqnarray}  
$V(H_{1},H_{2})$ contains the soft SUSY breaking terms:
\begin{eqnarray}
V(H_{1},H_{2})=\left(\left|\mu\right|^2 + m_{H_{1}}^{2}\right)\left|H_{1}\right|^{2}+\left(\left|\mu\right|^2 + m_{H_{2}}^{2}\right)\left|H_{2}\right|^{2}+b\left(\epsilon_{ab}H_{1}^{a}H_{2}^{b}+h.c.\right)\nonumber \\ 
~~+~~\frac{1}{2}g^{2}\left|H_{1}^{\dagger}H_{2}\right|^{2}+\frac{1}{8}\left(g^{2}+g'^{2}\right)\left(\left|H_{2}\right|^{2}-\left|H_{1}\right|^{2}\right)^{2}.\label{eq:Exp-H-Pot}
\end{eqnarray}
The terms proportional to the mixing parameter $\left|\mu\right|$ come from the $F$-contribution to the SUSY Lagrangian, the terms with the EW gauge couplings ($g$, $g'$) come from the $D$-contribution while $m_{H_{1}}^{2}$, $m_{H_{2}}^{2}$ and $b$ are the soft SUSY breaking parameters \cite{Dimopoulos, Girardello}. The potential can be rotated to a basis where the quadratic terms in the real and imaginary components of $H_j^0$ are diagonalized through the transformations 
\begin{eqnarray}
\left(\begin{array}{c}
\chi_{1}^{0}\\
\chi_{2}^{0}
\end{array}\right)=\frac{1}{\sqrt{2}}D^{\dagger}(\beta)\left(\begin{array}{c}
\pi^{0}\\
A
\end{array}\right), &  \left(\begin{array}{c}
\phi_{1}^{0}\\
\phi_{2}^{0}
\end{array}\right)=D^{\dagger}(\alpha)\left(\begin{array}{c}
H\\
h
\end{array}\right), &  D(\theta)=\left(\begin{array}{cc}
c_{\theta} & s_{\theta}\\
-s_{\theta} & c_{\theta}
\end{array}\right).\label{eq:Pi-A}
\end{eqnarray}
In this basis the Higgs sector has five physical Higgs bosons, three of them are neutral, the other two are charged. In lowest order these are the lightest ($h$) and heavy ($H$) CP-even Higgs bosons, the CP-odd Higgs boson ($A$), and the two charged Higgs bosons ($H^{\pm}$). On the other hand, we can rewrite the bare mass Lagrangian related to the neutral fields $H_j^0$ as 
\begin{eqnarray}
\left(\begin{array}{cc}
H & h\end{array}\right)D(\alpha)\left(\begin{array}{cc}
M_{Z_0}^{2}c_{\beta}^{2}+M_{A_0}^{2}s_{\beta}^{2}+\frac{T_{01}}{\sqrt{2}\: v_{1}} & -\left(M_{A_0}^{2}+M_{Z_0}^{2}\right)s_{\beta}c_{\beta}\\
-\left(M_{A_0}^{2}+M_{Z_0}^{2}\right)s_{\beta}c_{\beta} & M_{Z_0}^{2}s_{\beta}^{2}+M_{A_0}^{2}c_{\beta}^{2}+\frac{T_{02}}{\sqrt{2}\: v_{2}}
\end{array}\right)D^{\dagger}(\alpha)\left(\begin{array}{c}
H\\
h
\end{array}\right)\nonumber \\ 
~~+~~~\left(\begin{array}{cc}
\pi^{0} & A\end{array}\right)\left(\begin{array}{cc}
m_{\pi_0}^{2} & 0\\
0 & M_{A_0}^{2}+\frac{T_{01}}{\sqrt{2}\: v_{1}}c_{\beta}^{2}+\frac{T_{02}}{\sqrt{2}\: v_{2}}s_{\beta}^{2}
\end{array}\right)\left(\begin{array}{c}
\pi^{0}\\
A
\end{array}\right),\label{eq:Lag-Mass-AH}
\end{eqnarray}
where 
\begin{eqnarray}
M_{A_0}^{2}=b\left(cot\beta+tan\beta\right), & M_{Z_0}^{2}=\frac{1}{4}\left(g^{2}+g'^{2}\right)\left(v_{1}^{2}+v_{2}^{2}\right), & tan\beta=\frac{v_{2}}{v_{1}}, \label{eq:Param}
\end{eqnarray}
and $T_{0j}$ is the coefficient of the linear term in $\phi_{j}^{0}=ReH_{j}^{0}$. The angle $\alpha$ is a function of the parameters of the theory: 
\begin{eqnarray}
tan(2\alpha)=tan(2\beta) \frac{ M_{A_{0}}^{2} + M_{Z_{0}}^{2}}{M_{A_{0}}^{2}-M_{Z_{0}}^{2}} & ; &  -\frac{\pi}{2} < \alpha < 0 ~ .  
\end{eqnarray}
We have used the short notation $s_\theta=sin(\theta)$ and $c_\theta=cos(\theta)$. At tree level the Higgs sector can therefore be parametrized in terms of the Z gauge boson mass ($M_{Z_0}$), the mass of the CP-odd Higgs boson ($M_{A_0}$) and $tan\beta$, which is the ratio of the two vacuum expectation values. 
The masses of the five Higgs bosons particles $h$, $H$, $A$ and $H^{\pm}$ follow as predictions. \\ The dominant contribution of the three-loop corrections to $M_h$ comes from the QCD sector of the MSSM Lagrangian. Having in mind that, we have derived the necessary renormalization conditions by imposing the following procedure. \\

i)~We~have computed the correction in the EW gaugeless limit at order O($\alpha_t\alpha_s^2$). Consequently, all the terms proportional to $M_{Z_0}$ are disregarded. In this limit the tree-level relation $\alpha = \beta$ is satisfied. We have also used the non-light fermions limit (NLF) where all the fermion masses are put to zero except the mass of the top quark, $M_t$. \\

ii) The numerical dominant contributions due to the higher order corrections to the Higgs pole masses can be obtained in the limit of vanishing external momentum, $p^2 = 0$, where $p$ is the momentum transferred in the external lines of the self-energy corrections. Up to two-loop level the shifts in the mass $M_h$ due to the momentum dependence are below 1~GeV in all scenarios studied, the bulk of the two-loop corrections comes from the effective-potential effects which are of the order of $10~\rm{GeV}$ \cite{Sophia}. A well-behaved perturbative expansion will make this dependence even weaker at three-loop level. Motivated by these observations we have adopted the approximation of zero external momentum and therefore we have avoided dealing with the Higgs wave function renormalization and also with the renormalization of $tan\beta$. As the wave functions renormalization of the Higgs fields don't make contributions, it is enough to rewrite (\ref{eq:Lag-Mass-AH}) in terms of the renormalized parameters $M_A$ and $T_j$ according to 
\begin{eqnarray}
X_{0}\rightarrow X+\delta X &  ; &  \delta X=\sum_{l=1}^{3}\delta^{(l)}X\:\:\:; \:\:\: X = M_A^2,\: T_{1,2} \: . \label{eq:CMZ}
\end{eqnarray}
iii) We have defined the vevs of the Higgs fields ($v_1$, $v_2$) as the minima of the full effective potential. The condition $\left\langle \Omega\left|H^0_{1,2}\right|\Omega\right\rangle =0$ is satisfied order by order in the perturbative expansion of the one-point Green function implying
\begin{eqnarray}
T_{1,2}^{tree}=0, & \delta^{(l)}T_{1,2}=-T_{1,2}^{(l)},\label{eq:T-CT}
\end{eqnarray}
where $T_{j}^{(l)}$ is the $l-$loop Higgs tadpole contribution.\\ Taking in consideration the above restrictions the expressions for the renormalized three-loop corrections to the neutral CP-even Higgs bosons are reduced to
\begin{eqnarray}
\left.\widehat{\sum}\right._{\psi_{i}\psi_{j}}^{(3)}=\left.\sum\right._{\psi_{i}\psi_{j}}^{(3)}~-~\delta^{(3)}M_{\psi_{i}\psi_{j}}^{2} &  ; &  \psi_j = h,\: H. \label{eq:RC}
\end{eqnarray}
The letter sigma with a hat represents the renormalized Higgs self-energies at zero momenta, the letter sigma without a hat is the unrenormalized Higgs self-energies and the terms with delta are the counter-terms of the CP-even Higgs boson masses whose expressions are: 
\begin{eqnarray}
\delta^{(3)}M_{hh}^{2}~=~\delta^{(3)}M_{AA}^{2}-\left(\frac{h_{t}}{2M_{t}}s_{\beta}\right)\left(\frac{T_{h}^{(3)}}{\sqrt{2}}s_{4\beta}+\frac{T_{H}^{(3)}}{\sqrt{2}}s_{2\beta}^{2}\right), \label{eq:hh-CT}\\ 
\delta^{(3)}M_{HH}^{2}~=~\left(\frac{h_{t}}{2M_{t}}s_{\beta}\right)\left(\frac{T_{h}^{(3)}}{\sqrt{2}}s_{4\beta}-\frac{T_{H}^{(3)}}{\sqrt{2}}c_{2\beta}^{2}\right), \label{eq:HH-CT}\\ 
\delta^{(3)}M_{hH}^{2}~=~\left(\frac{h_{t}}{2M_{t}}s_{\beta}\right)\left(\frac{T_{h}^{(3)}}{\sqrt{2}}c_{4\beta}+\frac{T_{H}^{(3)}}{2\sqrt{2}}s_{4\beta}\right), \label{eq:hH-CT}
\end{eqnarray}
where $h_t$ is the top Yukawa coupling, 
\begin{eqnarray}
\left(\begin{array}{c}
T_{H}\\
T_{h}
\end{array}\right)=D\left(\beta\right)\left(\begin{array}{c}
T_{1}\\
T_{2}
\end{array}\right) & \rm{and} &\delta^{(3)}M_{AA}^2=Re\left[\left.\sum\right._{AA}^{(3)}\left(M_{A}^{2}\right)\right]. \label{CT-MA}
\end{eqnarray}
We have imposed an on-shell renormalization condition for the renormalized self-energy of the neutral CP-odd Higgs boson. Finally, the renormalized CP-even Higgs boson masses are obtained by determining the poles of the inverse propagator matrix
\begin{eqnarray}
\left(\Delta_{H}\right)^{-1}=-i\left(\begin{array}{cc}
p^{2}-M_{H}^2+\left.\widehat{\sum}\right._{HH} & \left.\widehat{\sum}\right._{hH}\\
\left.\widehat{\sum}\right._{hH} & p^{2}-M_{h}^2+\left.\widehat{\sum}\right._{hh}
\end{array}\right), \label{eq:inverse}
\end{eqnarray}
which is equivalent to solve the determinantal equation
\begin{eqnarray}
\left[p^{2}-M_{H}^2+\left.\widehat{\sum}\right._{HH}\right]\left[p^{2}-M_{h}^2+\left.\widehat{\sum}\right._{hh}\right]-\left[\left.\widehat{\sum}\right._{hH}\right]^{2}=0. \label{eq:det}
\end{eqnarray}   
The three-loop counter-terms derived from the mass Lagrangian (\ref{eq:Lag-Mass-AH}) in the above expressions are useful to cancel local UV divergences. However, the unrenormalized self-energies as well as the tadpoles of the neutral CP-even and CP-odd Higgs boson fields can contain also non-local divergences coming from a sub-loop in the three-loop diagrams. It is therefore necessary an additional sub-renormalization procedure to remove these infinities. The procedure consists in the inclusion of one and two-loop vacuum diagrams with counter-term insertions which cancel those divergences arising in a sub-loop. At order O($\alpha_t\alpha_s^2$) we need the O($\alpha_s$)-contribution of the one-loop counter-terms coming from the renormalization of the gluino mass, the top quark mass, the squark masses and the stop mixing angles. In addition, we need the two-loop renormalization of the top mass, the stop masses and stop mixing angles at order O($\alpha_s^2$). We have got all the one and two-loop counter-terms in the $\overline{DR}$ scheme, where the UV divergences are minimally subtracted. In order to preserve supersymmetry to all perturbative orders we have used the regularization procedure DRED \cite{DRED}. In Appendix~\ref{ap:non-local} we describe the main renormalization conditions to remove the sub-divergences from the three-loop diagrams and we give the expressions of the counter-terms.

\section{\large Technical Details to the Three-Loop Calculation of $\rm{M_h}$}
\label{sec-2}

\begin{figure}
\centering
\includegraphics[width=35pc]{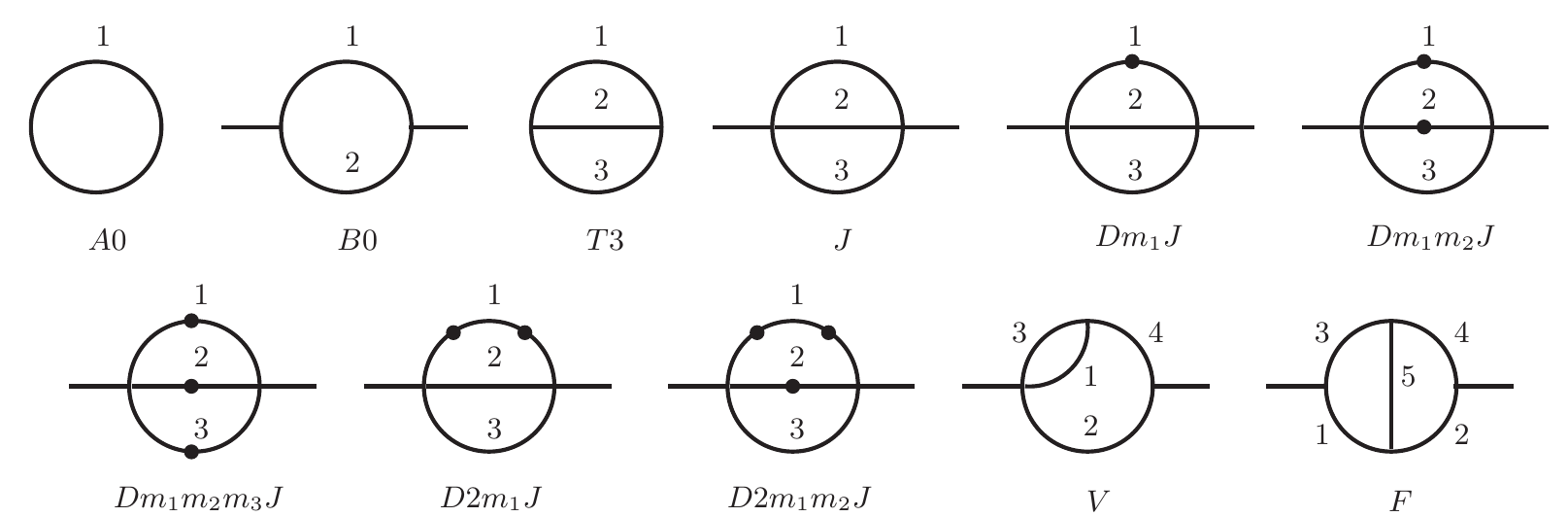}\hspace{2pc}%
\caption{\small{Basis of one and two -loop master integrals obtained from the Tarasov method. The dot over the internal line $j$ represents a partial derivative of the propagator regarding the squared mass $m_j^2$.}}
\label{MI2}
\end{figure}

In this section we are going to discuss the technical details of our three-loop diagrammatic computation. We restrict the calculations to the SUSY-QCD sector where the Higgs couples just to the top quark or its super-partner. That is, we are interested in the terms of order $\alpha_{s}^{2}\alpha_t$ in the perturbative expansion of the Higgs mass. The three-loop radiative correction to $M_h$ is obtained by evaluating the neutral Higgs boson self-energies ($\sum$) and the tadpoles ($T$) for the fields $h$, $H$ and $A$ according with the equations (\ref{eq:RC}) and (\ref{eq:hh-CT}). All the needed Feynman diagrams and their corresponding amplitudes are generated with the package FeynArts \cite{FeynArts}. FeynArts generates the amplitudes without taking the contractions of the Dirac and Color indices. As a result, we obtain a total set of $7738$ self-energies and $7180$ tadpoles which are moreover not regularized. With the help of the Mathematica package FeynCalc \cite{FeynCalc} we have written a routine that implements the regularization of Feynman integrals by dimensional reduction. In this scheme all gamma-matrices are of the quasi-four-dimensional space (Q4S) \cite{Dominik, Gnedinger, Dominik2} while loop-momenta remain in the quasi-D-dimensional space. Thus, we have performed the Dirac and Lorentz algebra over the numerators of the integrals using the Q4S algebra. It is worth to mention that the Q4S algebra for diagrams which contain the cubic vertex quark-squark-gluino requires a special treatment. This vertex introduces the $\gamma_5$ matrix in the Dirac traces. We have dealt $\gamma_5$ as an anti-commuting object, which is allowed in our computation because traces with an odd number of $\gamma_5$ always contain less than four gamma matrices \cite{Jegerlehner}. By other side, sum over the Color indices was done with the help of the package SUNSimplify of FeynCalc. After performing the Dirac and the Color algebra, each amplitude can be expressed as a superposition of a set of "scalar integrals" which can contain propagators with negative powers (irreducible numerators). We need in our calculation a total of $3525$ of such integrals. We have avoided the use of asymptotic expansions at the integral level in order to obtain a general expression valid for any election of the input masses. For this reason we have exploited the fact that this set of integrals are not independent of each other but related by the integration by parts (IBP) and Lorentz invariance (LI) relations. We have used the IBPs to generate a homogeneous system of linear equations where the scalar integrals are the unknowns. The system can be reduced to a small set of irreducible integrals, the so called Master Integrals. This is something that cannot be done by hand because there are thousand of equations. Thus, we have used the program Reduze~\cite{Reduze}, an implementation of the Laporta algorithm, in order to carry out this reduction. 
\begin{figure}
\centering
\includegraphics[width=35pc]{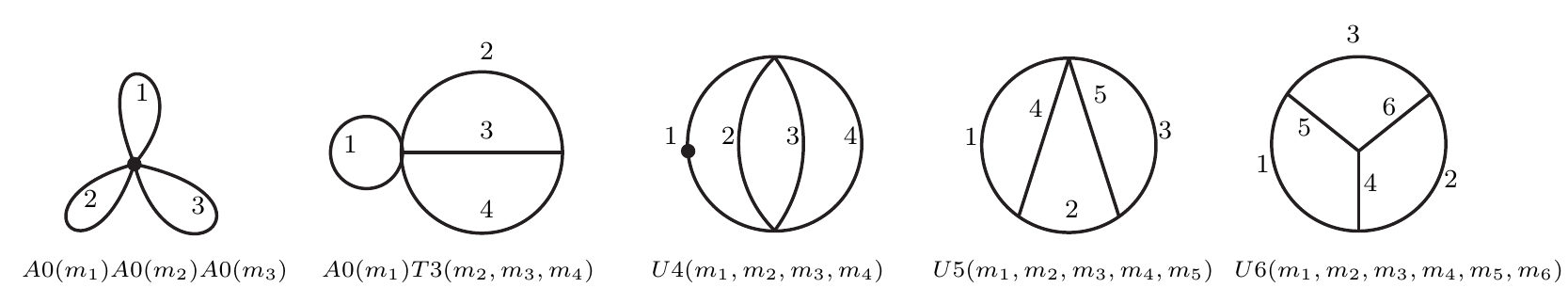}\hspace{2pc}%
\caption{\small{Basis of three-loop vacuum integrals obtained from the IBP reduction with Reduze.}}
\label{MI}
\end{figure}
We have found a basis of 32 master integrals which correspond to different mass configurations for the five vacuum diagrams depicted in Figure~\ref{MI}, where each topology can contain at most four independent mass scales. The one-loop function $A0$ and the two-loop function $T3$ have very well known analytical expressions for any configuration of masses, while the three-loop functions $U4$, $U5$ and $U6$ have an analytical solution just for the cases where there are one or two independent mass scales but when there are three or four independent scales a numerical evaluation is required. The analysis of convergence and the numerical evaluation of the three-loop master integrals which are unknown analytically were performed with the help of the program TVID developed by A. Freitas \cite{Freitas2}. TVID uses the discontinuities coming from the one-loop self-energy ($B0$) and the one-loop vertex ($C0$) to produce dispersion relations that are useful in the evaluation of the three-loop vacuum integrals in terms of one and two -dimensional integral representations. In all cases we have been able to get analytical expressions for the divergent part while the finite part could be numerically evaluated with up to 10 digits of accuracy. It is possible to reach this precision because the numerical integrations are at most 2-dimensional and therefore there is a controlled treatment of any singularities (see Appendix~\ref{ap:dispersion}). \\ To sum up, each three-loop amplitude can be expressed as a linear combination of the integrals drawn in Figure~\ref{MI} with coefficients that are ratios of polynomials in the masses and the space-time dimension. These coefficients can contain poles of first and second order and therefore the renormalized correction to $M_h$ requires also the evaluation of the evanescent terms of the master integrals up to second order, that is to say, the terms at order O($\varepsilon$) and O($\varepsilon^2$) in the Laurent expansion. It is no possible to evaluate these contributions with TVID because the higher-$\varepsilon$ terms of the real and imaginary parts of $B0$ and $C0$ and therefore of $U4$, $U5$ and $U6$ are not included in the program. For this reason we have used the code SecDec \cite{SecDec} which admit a numerical evaluation of the evanescent terms. \\
By other side, we also need to generate the amplitudes for the diagrams which are responsible for removing the non-local sub-divergences. We have written a routine in Mathematica that generates all the expressions for the counter-terms listed in Appendix~\ref{ap:non-local}. The generation of the involved regularized amplitudes was done with the help of the FeynArts and FeynCalc functions. In contrast to the three-loop diagrams, the one and two-loop counter-terms are determined from the evaluation of fermionic and scalar self-energies with the external momentum transferred different from zero, $p^2 \neq 0$. The resulting self-energies can be further reduced using the Tarasov method \cite{Tarasov}, that is implemented in the code TARCER \cite{TARCER}, to the basis of one and two -loop master integrals represented by the diagrams of Figure~\ref{MI2}. The one-loop counter-terms of the top quark mass, the gluino mass, the sfermion masses and the stop mixing angles can be determined in terms of the Passarino-Veltman functions~$A0$~and~$B0$. The two-loop counter-terms of the stop masses and mixing angles can be expressed as a superposition of the eleven master integrals depicted in Figure~\ref{MI2}. 
 \\
In the $\overline{DR}$ scheme we just need the divergent parts of the integrals. The one-loop functions $A0$ and $B0$ have the analytical expressions found in Appendix~\ref{ap:ren-const}. The functions $Dm_1 m_2 J$, $Dm_1 m_2 m_3 J$ and $D2m_1 m_2 J$ are finite. $D2m_1J$ has only a 1/$\varepsilon$ divergence with coefficient $-1/2m_{1}^{2}$, independent of the values of $m_2$ and $m_3$. $Dm_1J$ has poles of one and second order, its 1/$\varepsilon^2$ divergence is 1/2, mass independent while its 1/$\varepsilon$ divergence is $1/2-ln(m_1^2/\mu_r^2)$. The 1/$\varepsilon^2$ divergence coefficient of $J$ is $1/2(m_1^2+m_2^2+m_3^2)$ while the 1/$\varepsilon$ divergence coefficient is
\begin{eqnarray}
\frac{3}{2}\left(m_{1}^{2}+m_{2}^{2}+m_{3}^{2}\right)-\frac{p^{2}}{4}-\sum_{j=1}^{3}m_{j}^{2}ln\left(m_{j}^{2}/\mu_r^2 \right) \label{eq:divcoeff}.
\end{eqnarray} 
For the cases where one or more masses $m_j^2$ vanish in the expression (\ref{eq:divcoeff}) one should take the zero order term of the Taylor expansion around $m_j^2=0$ to get the right expression. The same apply for $T3$ but choosing the external momentum $p^2$ equal to zero. Finally, we need the functions $F$ and $V$ with at most three independent mass scales. The function $F$ is finite, while $V\left[p^{2},m_{1}^{2},m_{2}^{2},m_{3}^{2},m_{4}^{2}\right]$ have a divergence 1/$\varepsilon^2$ with coefficient 1/2 and a 1/$\varepsilon$ divergence equal to $B0_{fin}\left[p^{2},m_{2}^{2},m_{4}^{2}\right]+1/2$, where $B0_{fin}$ refers to the finite part of the function $B0$. All these analytical expressions were numerically checked with the code SecDec, where the prefactor $Exp[(-2\gamma_E\varepsilon)]$ must be specified in order to get the correct result. In Appendix~\ref{ap:ren-const} the counter-term expressions involved in the renormalization of the non-local ultraviolet divergences are listed. These results can be checked with those of the review \cite{Luminita} and references therein.

\section{\large Results and Numerical Analysis}
\label{sec-3}

\begin{figure}
\centering
\subfigure[\footnotesize{$\mu_r=245.7~\rm{GeV}$, $tan\beta=10$, $M_A=1~\rm{TeV}$}]{
\includegraphics[width=18pc]{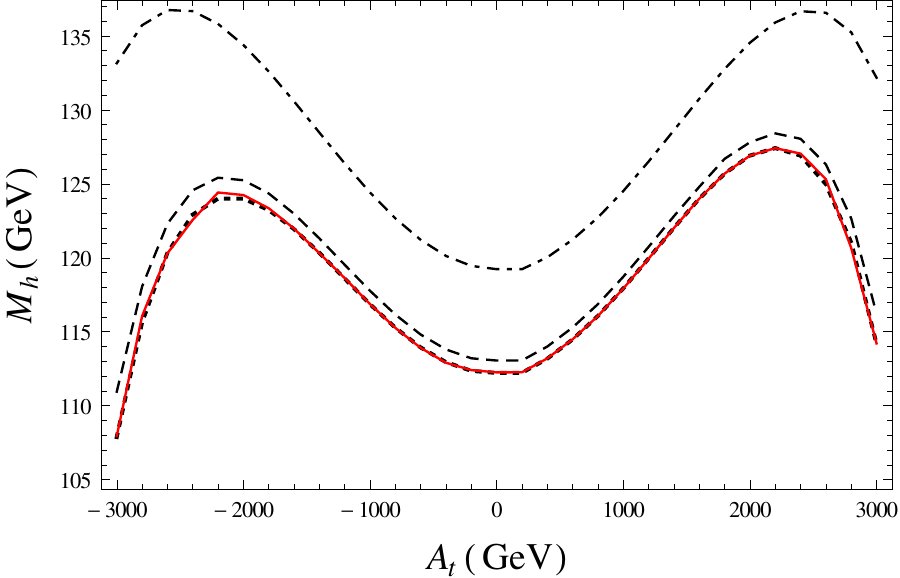} \label{fig:At}}
\subfigure[\footnotesize{$\mu_r=245.7~\rm{GeV}$, $A_t=1,7~\rm{TeV}$, $M_A=1~\rm{TeV}$}]{
\includegraphics[width=18pc]{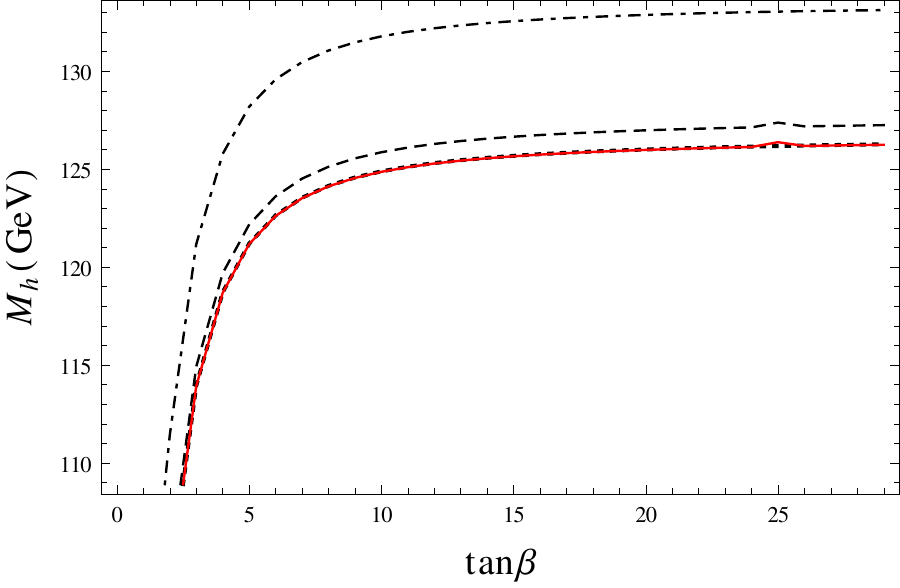} \label{fig:Tb}}
\caption{\small{Dependence of $M_h$ on (a) $A_t$ and (b) $tan\beta$. The dot-dashed and dashed lines are the one and two -loop predictions of FeynHiggs. The dotted line is the three-loop prediction of H3m and the red solid line depicts our three-loop predictions.}}
\label{NumA}
\end{figure}
\begin{figure}
\centering
\subfigure[\footnotesize{$A_t=1,7~\rm{TeV}$, $M_A=1~\rm{TeV}$}]{
\includegraphics[width=18pc]{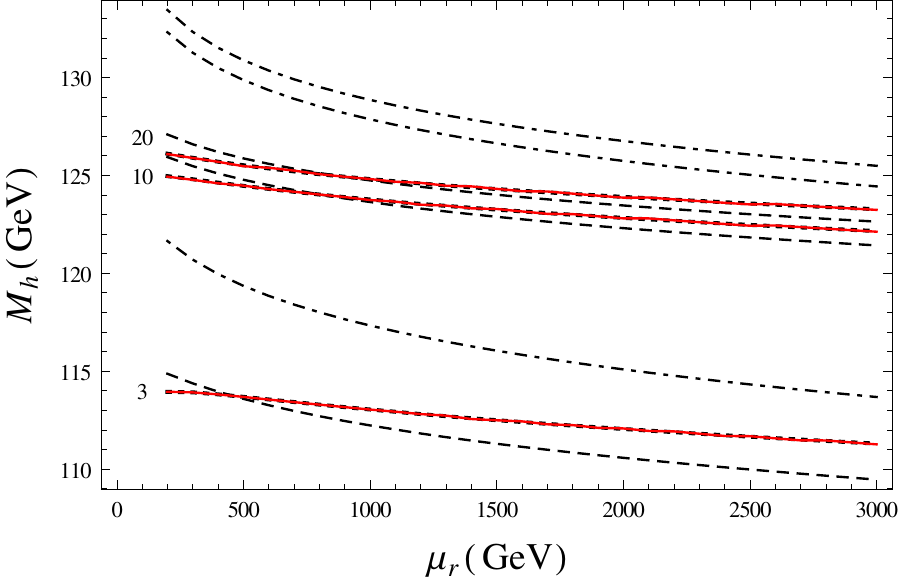} \label{fig:mu}}
\subfigure[\footnotesize{$tan\beta=10$, $M_A=1~\rm{TeV}$}]{
\includegraphics[width=18pc]{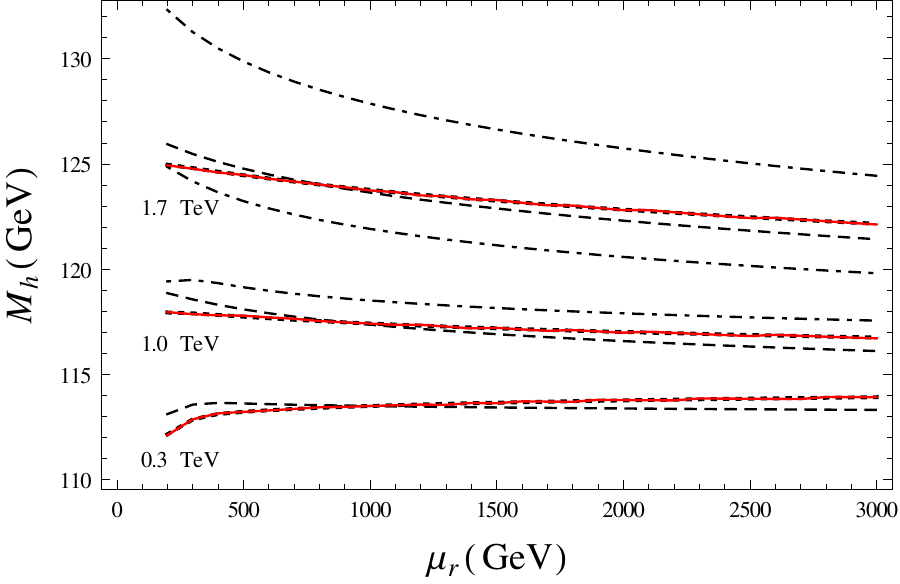} \label{fig:mu2}}
\caption{\small{Dependence of $M_h$ on $\mu_r$. (a) The evolution for three different values of $tan\beta$ is studied. We consider $tan\beta$ = 3, 10 and 20. (b) $M_h$ is plotted for the scenarios with $A_t$ = 0.3 TeV, 1 TeV and 1.7 TeV. The inclusion of three-loop corrections reduces the scale dependence by a factor between 1.5 - 2.0. The scale dependence is improved when we consider lower values of $tan\beta$ and $A_t$. To draw these plots we have used the same conventions as in  the Figure \ref{NumA}.}}
\label{NumB}
\vspace*{0.5cm}
\end{figure}

Once the local and non-local UV divergences have been subtracted from the Higgs self-energies, we get finite three-loop corrections to the CP-even and CP-odd Higgs boson masses that are useful in the derivation of the renormalized value of $M_h$ which is obtained as a solution of the pole equation (\ref{eq:det}). Our corrections depend upon 26 parameters: the renormalization scale $\mu_{r}$, the SM $\overline{DR}$ parameters $h_t$, $M_t$, $\alpha_s$ and the MSSM parameters $\mu$, $tan\beta$, $M_A$, $M_{\tilde{g}}$, $\theta_t$, $\tilde{m}_{f_{1,2}}$ and $A_f$, with $f=u,d,t,b,c,s$. Their values as well as the renormalization group evolution of the SM parameters are determined with the help of the spectrum generator SoftSUSY~\cite{SoftSUSY}. We use the package SLAM~\cite{SLAM} (Supersymmetry Les Houches Accord with Mathematica) in order to export to Mathematica any needed parameter generated with SoftSUSY. \\ We have performed a numerical comparison between our three-loop predictions and the other higher order corrections currently included in FeynHiggs \cite{FeynHiggs, FeynHiggs2} and the three-loop results implemented in H3m \cite{Harlander1, Harlander2} combined with the lower-order results of FeynHiggs. 
We discuss our results in three different limits: i) the $m_h^{free}$ scenario, where $\mu_{r}$, $tan\beta$, $M_A$ and $A_t$ are left as free input parameters. ii) The $m_h^{max}$ and $m_h^{mod+}$ scenarios analyzed in \cite{Carena2}. In these three different scenarios we don't make any specific assumptions about the soft SUSY-breaking mechanism and we interpret the LHC signal at 125~GeV as the lightest CP-even Higgs boson. We consider values of the SUSY scale in the region $M_{SUSY}<1.2~\rm{TeV}$, where the combined theoretical uncertainty of the fixed-order calculation is lesser than the combined uncertainty of the effective field theory (EFT) calculation~\cite{Allanach}. At the critical point $M_{SUSY}=1.2~\rm{TeV}$ the fixed-order and EFT combined uncertainties are equal and a hybrid calculation should be used \cite{Bahl1,Bahl2,Bahl3}. Above the critical point the EFT computation is more accurate  and therefore an EFT approach, where an effective SM is used below a super-symmetric scale \cite{Villadoro,Draper,Bagnaschi}, should be preferred. In this paper we discuss our numerical results in scenarios where a fixed-order calculation is recommended. \\
In the $m_h^{free}$ scenario we fix the parameters $M_{\tilde{t}_L}=M_{\tilde{t}_R}=\tilde{m}_{q_{1,2}}=M_{SUSY}=1~\rm{TeV}$, where $q$ denotes any quark different than the top quark. We also set $M_{\tilde{g}}=1500~\rm{GeV}$, $\mu=200~\rm{GeV}$ and $M_A = 1000~\rm{GeV}$. Using this limit we have studied the dependence of the Higgs boson mass $M_h$ on the soft-breaking parameter $A_t$ (Fig-\ref{fig:At}) on the input parameter $tan\beta$ (Fig-\ref{fig:Tb}) and on the renormalization scale $\mu_r$ (Fig-\ref{fig:mu} and Fig-\ref{fig:mu2}). \\ 
At one and two -loop level we have generated the Higgs mass predictions  with the help of the code FeynHiggs. These contributions are represented in the plots with the dot-dashed and dashed curves respectively. This convention has been used in all panels. We have included the full one and two-loop corrections ($\alpha_s\alpha_t$, $\alpha_s\alpha_b$, $\alpha_t^2$, $\alpha_t\alpha_b$, $\alpha_b^2$) in the rMSSM. The one-loop field-renormalization constants and the one-loop $tan\beta$ counter-term are set in the $\overline{DR}$ scheme. We don't assume any approximation for the external momentum value for the one and two-loop corrections, i.e. we set in FeynHiggs a full determination of the propagator matrix poles (\ref{eq:inverse}). Besides, the dotted curve represents the three-loop predictions at order $O(M_t^2\alpha_t\alpha_s^2)$ coming from the program H3m while the red solid line represents our three-loop prediction evaluated at the same order. Our three-loop results shown in Figure~\ref{NumA} are quite sizeable, amounting a size between $0.8$~to~$3.1$~GeV compared to the two-loop corrections and $-0.373$ to $0.418$~GeV regarding the three-loop prediction of H3m. The relative size and sign of the corrections depend on our election of the renormalization scheme. 
Figure~\ref{fig:Tb} shows a strong dependence on $tan\beta$ for small values close to $tan\beta = 3$, while for large values above $tan\beta = 10$ the variation of $M_h$ is marginal and closer to the LHC Higgs mass value. Figure~\ref{NumB} depicts the dependence of $M_h$ on the renormalization scale $\mu_r$. In Fig-\ref{fig:mu} the dependence is studied for three different values of $tan\beta$, namely $tan\beta=3,10$ and $20$. The three-loop corrections lead to a more stable dependence of $M_h$ with the renormalization scale $\mu_r$ than the one and two -loop predictions, reducing the scale dependence by a factor between 1.5 and 2.0. This stability increases for lower values of $tan\beta$ in the $m_h^{free}$ scenario. Fig-\ref{fig:mu2} shows the RG evolution for three values of soft breaking parameter $A_t$, $A_t=0.3~\rm{TeV},1~\rm{TeV}$ and $1.7~\rm{TeV}$. The evolution is more stable when the three-loop corrections are added and when the value of $A_t$ decreases, reducing the scale dependence by a factor of about 1.6 compared to the two-loop predictions.
\begin{table}
\centering
\begin{tabular}{|c|c|c|c|c|c|}
\cline{2-6} 
\multicolumn{1}{c|}{} & $M_{t}$ & $M_{SUSY}$ & $X_{t}$ & $M_{\tilde{g}}$ & $\mu$\tabularnewline
\hline 
$m_{h}^{max}$ & 173.2 GeV & 1000 GeV & 2$M_{SUSY}$ & 1500 GeV & 200 GeV\tabularnewline
\hline 
$m_{h}^{mod+}$ & 173.2 GeV & 1000 GeV & +1.5$M_{SUSY}$ & 1500 GeV & 200 GeV\tabularnewline
\hline 
\end{tabular}
\caption{Input parameters for the $m_{h}^{max}$ and $m_{h}^{mod+}$ scenarios.}
\label{tb:table1}
\end{table}
In the $m_h^{max}$ and $m_h^{mod+}$ scenarios the renormalization scale is set to $\mu_r=m_t=173.2~\rm{GeV}$, where $m_t$ represents the combined Tevatron and LHC experimental value of the top quark mass~\cite{tmass}. Besides one has to fix $M_{\tilde{t}_L}=M_{\tilde{t}_R}=M_{SUSY}$, $A_q=0$ and $\tilde{m}_{q_{1,2}}=1500$~GeV. The parameter $A_t$ is fixed through the mixing term in the squark sector, $X_t = A_t - \mu cot\beta$, while $M_A$ and $tan\beta$ are left as free parameters. Within the $m_h^{max}$ scenario $X_t$ is chosen in order to maximize the value of $M_h$ for a given election of $\mu$ and $\tan\beta$. This occurs when $\left| X_t/M_{SUSY}\right|\approx 2$, where the radiative corrections in the FD calculation reach to the largest positive contribution. With this election of parameters the mass of the lightest CP-even Higgs boson is in agreement with the LHC Higgs boson signal just in a relatively small strip in the $M_A$-$tan\beta$ plane. A convenient way to enlarge the region of validity is to decrease the amount of mixing in the stop sector. The $m_h^{mod+}$ scenario is a modification of $m_h^{max}$ where this mixing represented by $\left| X_t/M_{SUSY}\right|$ is reduced. In detail we consider the input parameters shown in~Table~\ref{tb:table1} for each scenario. The stop mixing angle $\theta_t$ and the stop masses $\tilde{m}_{t_{1}}$, $\tilde{m}_{t_{2}}$ are functions of the parameters specified in~Table~\ref{tb:table1}. 
\begin{figure}[h]
\centering
\vspace*{1cm}
\subfigure[\footnotesize{$\mu_r=173.2~\rm{GeV}$}]{
\includegraphics[width=18pc]{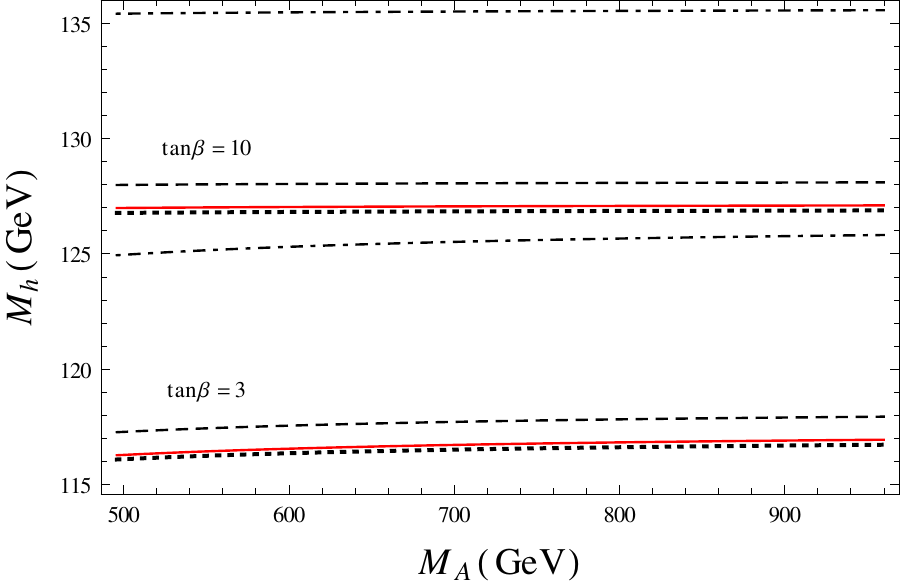} \label{fig:MAmax}}
\subfigure[\footnotesize{$\mu_r=173.2~\rm{GeV}$, $tan\beta=3$ }]{
\includegraphics[width=18pc]{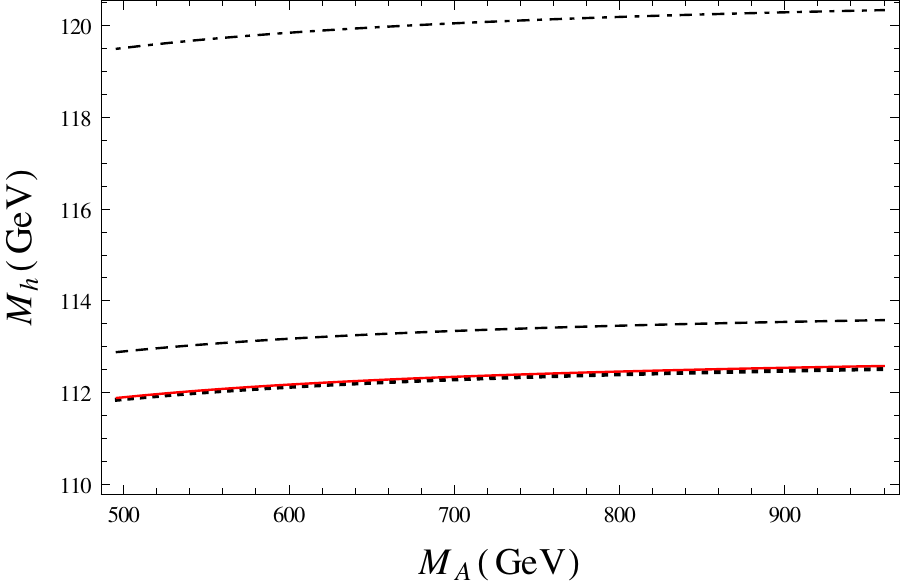} \label{fig:MAmod}}
\caption{\small{The lightest Higgs boson mass $M_h$ as a function of $M_A$ in (a) the $m_h^{max}$ limit for $tan\beta=3, 10$ and (b) in the $m_h^{mod+}$ scenario. Our three-loop corrections are, within theoretical uncertainties, in agreement with the predictions of H3m. The difference can amount a size of about 0.2 GeV in $m_h^{max}$ and 0.05 GeV in $m_h^{mod+}$. These plots follow the same conventions as in the Figure \ref{NumA}.}}
\label{NumC}
\vspace*{0.5cm}
\end{figure}\\
To draw the plots in Figure~\ref{NumC} the~CP-odd Higgs boson mass $M_A$ is varied in the interval: $500~\rm{GeV}\leq M_A \leq 1~\rm{TeV}$ and we consider the values of $tan\beta=3$ and $tan\beta=10$. The numerical values of our three~loop corrections to $M_h$ are reduced compared with the two-loop predictions showing a good behaviour of the perturbative expansion. In both scenarios the three-loop corrections give rise to a significant reduction of about $1~\rm{GeV}$ to the two-loop value of the Higgs boson mass. In addition the values of the corrections are consistent with the results obtained with H3m. The difference can amount a size of about 0.2 GeV in the $m_h^{max}$ limit and 0.05 GeV in $m_h^{mod+}$. The detailed analysis of the individual sources of uncertainty of the three-loop fixed order $\overline{DR}\:'$ Higgs boson mass prediction of SoftSUSY, developed recently in \cite{Allanach}, shows a combined uncertainty for a maximal stop mixing scenario who varies between 1-4 GeV depending on the SUSY scale. Our results have then very similar numerical values to the ones obtained with H3m with differences which are within the combined theoretical uncertainty.         	     

\section{Conclusions}
\label{sec-4}
Following the Feynman diagrammatic procedure we have obtained a finite expression for the renormalized three-loop corrections to the neutral CP-even and CP-odd Higgs boson masses at the fixed-order $O(\alpha_t\alpha_s^2)$ in the rMSSM. We have computed only the dominant contributions coming from the gauge-less and the non-light-fermions limits and the approximation of vanishing external momentum transferred in the calculation of the involved self-energy functions. We have reduced the three-loop corrections, using the integration-by-parts recurrence relations, to a set of five master integrals with at most four independent mass scales whose numerical evaluation is possible thanks to the dispersion relation techniques. The three-loop vacuum integrals are expressed in terms of one and two-dimensional numerical integrals of elementary functions which can be efficiently evaluated for a general mass pattern. In this way we avoided the asymptotic expansions on the amplitudes at the integral level in the realistic mass hierarchies proposed in~\cite{Harlander2}. Thus, our work represents an independent check of the results implemented in H3m but also provides a numerical evaluation of the three-loop corrections in the whole supersymmetric parameter space without any assumption about the mass hierarchy. \\
We have studied the numerical impact of our three-loop corrections in the value of the pole mass $M_h$ for three different benchmark limits: the $m_h^{free}$, $m_h^{max}$ and $m_h^{mod+}$ scenarios. We have considered scenarios with the SUSY scale $M_{SUSY}$ lower than $1.2~\rm{TeV}$, where a fixed-order calculation have a combined theoretical uncertainty better than the estimated uncertainty from an EFT calculation~\cite{Allanach}. We have investigated the dependence of $M_h$ on $tan\beta$, the soft-breaking parameter $A_t$, the renormalization scale $\mu_r$ and the CP-odd mass $M_A$. The three-loop corrections show a good perturbative behaviour, their numerical size is about ten times lesser that the two-loop predictions and the dependence on the renormalization scale is reduced by around a factor two. The contributions yield a shift in the value of the lightest Higgs boson mass of the order of $1~\rm{GeV}$ compared with the two-loop results of FeynHiggs and a shift between $ -0.4~\rm{and}~0.4 ~\rm{GeV}$ compared with the three-loop predictions of H3m combined with the lower-order contributions of FeynHiggs. Our results are in agreement within the theoretical composited uncertainties with the predictions of $M_h$ obtained from H3m if the same renormalization scheme and kinematical limits of the SUSY-QCD mass spectrum are employed. \\ 
The corrections presented in this paper were computed without taking any assumption about the soft SUSY-breaking mechanism so that a direct comparison with the H3m results in the MSUGRA scenarios requires a reparametrization of our expressions and therefore is beyond the scope of this article. Moreover, there is still missing a comparison with the other three-loop contributions currently found in literature, namely the pure $\overline{DR}$ three-loop results recently presented in \cite{Himalaya} and implemented in FlexibleSUSY+Himalaya, the EFT calculations implemented in HSSUSY \cite{Athron}, the 2-loop + O($\alpha_t\alpha_s^2$)~3-loop results as implemented in FeynHiggs via log-resum and also the FeynHiggs full corrections including all logs resummed \cite{FeynHiggs2}. We left the numerical analysis of these missing comparisons as the subject of future works. \\

\textbf{Acknowledgements.} We would like to thank to Dr. S. Borowka for very useful discussions about the technical aspects of our three-loop computation. We thank to group of Prof. Dr. B. A. Kniehl for their hospitality and financial support during the stay of E. A. Reyes R. in DESY. Our computation was done using the facilities of the II. Institute for Theoretical Physics, Hamburg University. A. R. Fazio acknowledges the support received by the Erwin Schr\"{o}edinger International Institute for Mathematical Physics, where the final stages of this work have been completed. This work is partially financially supported by the research grant N. 34729, Minimal Supersymmetric Standard Model Higgs Boson to High Accuracy of the call \small{CONVOCATORIA NACIONAL DE PROYECTOS PARA EL FORTALECIMIENTO DE LA INVESTIGACION, CREACION E INNOVACION DE LA UNIVERSIDAD NACIONAL DE COLOMBIA 2016-2018}.

\newpage

\section*{APPENDICES}

\vspace*{1cm}

\appendix

\section{\large Dispersion Method for Three-Loop Vacuum Integrals}
\label{ap:dispersion}

The three-loop vacuum integrals are an important block in the determination of the three-loop corrections to the Higgs boson mass as was discussed in Section \ref{sec-2}. Besides integrals that factorize into products of one and two-loop contributions, the integrals U4, U5 and U6 defined as
\begin{eqnarray}
U4\left(m_{1},m_{2},m_{3},m_{4}\right)\equiv {\rm INT}\left[2,1,1,1,0,0\right], ~~~~~~~~~~~\\ 
U5\left(m_{1},m_{2},m_{3},m_{4},m_{5}\right)\equiv { \rm INT}\left[1,1,1,1,1,0\right],~~~~~\\ 
U6\left(m_{1},m_{2},m_{3},m_{4},m_{5},m_{6}\right)\equiv {\rm INT}\left[1,1,1,1,1,1\right],
\end{eqnarray}
where
\begin{eqnarray}
{\rm INT}\left[\nu_{1},\nu_{2},\nu_{3},\nu_{4},\nu_{5},\nu_{6}\right]=i\frac{e^{3\gamma_{E}\varepsilon}}{\pi^{3D/2}}\int\prod_{j=1}^{3}d^{D}q_{j}\frac{1}{\left[q_{1}^{2}-m_{1}^{2}\right]^{\nu_{1}}\left[\left(q_{1}-q_{2}\right)^{2}-m_{2}^{2}\right]^{\nu_{2}}}\times \nonumber \\ 
\frac{1}{\left[\left(q_{2}-q_{3}\right)^{2}-m_{3}^{2}\right]^{\nu_{3}}\left[q_{3}^{2}-m_{4}^{2}\right]^{\nu_{4}}\left[q_{2}^{2}-m_{5}^{2}\right]^{\nu_{5}}\left[\left(q_{1}-q_{3}\right)^{2}-m_{6}^{2}\right]^{\nu_{6}}},
\end{eqnarray}
constitute a basis of master integrals for the three-loop corrections to the neutral CP-even and CP-odd Higgs boson self-energies in the approximation of vanishing external momentum transferred. The indices $\nu_j$ are integers numbers, $\varepsilon = (4-D)/2$ and $D$ is the number of the space-time dimensions. In this work we have used a method based on dispersion relations \cite{Freitas1} in order to get a numerical integration of the finite part of the  three-loop vacuum integrals. Using the dispersion relation techniques the four-propagator function $U4$ and the five-propagator function $U5$ can be represented in terms of one-dimensional integrals. In analogous way the six-propagator function $U6$ can be represented by a two-dimensional numerical integral. For the case of the $U4$ function the representation is
\begin{eqnarray}
U4\left(m_{1},m_{2},m_{3},m_{4}\right)=-\frac{e^{\gamma_{E}\varepsilon}}{i\pi^{D/2}}\int d^{D}q_{3}\int_{0}^{\infty}ds\frac{\Delta I_{db}\left(s\right)}{q_{3}^{2}-s+i\varepsilon} \nonumber \\  ~~
=~U4_{div}-\int_{0}^{\infty}ds\:A0_{fin}\left(s\right)\:\Delta I_{db,fin}\left(s\right),
\end{eqnarray}
where
\begin{eqnarray}
\Delta I_{db,fin}(s,m_1^2,m_2^2,m_3^2,m_4^2)=\Delta B_{0,m_{1}}\left(s,m_{1}^{2},m_{2}^{2}\right){\rm Re}\left[B0\left(s,m_{3}^{2},m_{4}^{2}\right)-B0\left(s,0,0\right)\right] \nonumber \\
~-~\Delta B_{0,m_{1}}\left(s,m_{1}^{2},0\right){\rm Re}\left[B0\left(s,m_{3}^{2},0\right)+B0\left(s,m_{4}^{2},0\right)-2B0\left(s,0,0\right)\right] \nonumber \\ 
~+~{\rm Re}\left[B_{0,m_{1}}\left(s,m_{1}^{2},m_{2}^{2}\right)\right]\left(\Delta B0\left(s,m_{3}^{2},m_{4}^{2}\right)-\Delta B0\left(s,0,0\right)\right) \nonumber \\ 
~-~{\rm Re}\left[B_{0,m_{1}}\left(s,m_{1}^{2},0\right)\right]\left(\Delta B0\left(s,m_{3}^{2},0\right)+\Delta B0\left(s,m_{4}^{2},0\right)-2\Delta B0\left(s,0,0\right)\right).
\end{eqnarray}
$U4_{div}$ contains the UV divergences of $U4$ while $\Delta B0$ and $\Delta B_{0,m_{j}}$ are the discontinuities of the scalar one-loop self-energy function, $B0$, and its mass derivative, $B_{0,m_{j}}=\frac{\partial}{\partial m_j^2} B0$, given by
\begin{eqnarray}
\Delta B0\left(s,m_{a}^{2},m_{b}^{2}\right)=\frac{1}{s}\lambda\left(s,m_{a}^{2},m_{b}^{2}\right)\Theta\left(s-\left(m_{a}+m_{b}\right)^{2}\right),\\ 
\Delta B_{0,m_{1}}\left(s,m_{a}^{2},m_{b}^{2}\right)=\frac{m_{a}^{2}-m_{b}^{2}-s}{s\lambda\left(s,m_{a}^{2},m_{b}^{2}\right)}\Theta\left(s-\left(m_{a}+m_{b}\right)^{2}\right).
\end{eqnarray}
Here $\lambda(x,y,z)$ is the K\"{a}llen function defined as 
\begin{equation}
\lambda(x,y,z)=\sqrt{x^2+y^2+z^2-2(xy+yz+zx)} \label{kallen}
\end{equation}
and $\Theta$ is the Heaviside step function. The finite part of the vacuum integral $U4$ is therefore expressed as a numerical integral of a combination of elementary functions, such a logarithms and square roots, which can be efficiently evaluated with numerical methods for general mass patterns without make any assumptions about the mass hierarchy of the SUSY particles. Analogous but more complicated expressions can be derived for the functions $U5$ and $U6$ and can be consulted in detail in \cite{Freitas1, Freitas2}. The implementation of these integrals can be found in the file U45.cc of the public code TVID-Version-1.1 (http://www.pitt.edu/~afreitas/). Internally the program uses the Gauss-Kronrod routine QAG from the Quadpack library \cite{Quadpack} to evaluate the dispersion integrals. This routine has been amended to facilitate 30 digit floating point arithmetic from the package doubledouble \cite{doubledouble} allowing results with at least ten digits of precision for the $U4$ and $U5$ integrals and a precision of eight digits for the $U6$ vacuum integral.

\section{\large Renormalization of Non-Local Divergences}
\label{ap:non-local}
This appendix presents analytic expressions for the counter-terms needed in the sub-renormalization procedure discussed in Section \ref{sec-2}. We have adopted a quite simple renormalization of the fermion masses, the gluino and the top quark masses are renormalized at one-loop in the $\overline{DR}$ scheme according with the condition:
\begin{eqnarray}
\frac{\delta^{(1)} M_{f}}{M_{f}}={\rm Dp} \left[\Sigma_{f}^{V}\left(M_{f}^{2}\right)+\Sigma_{f}^{S}\left(M_{f}^{2}\right)\right], \label{eq:mf-ct} 
\end{eqnarray}
where the fermion self-energy was decomposed into a vector, an axial-vector, a scalar and a pseudo-scalar part as
\begin{eqnarray}
\Sigma_{f}\left(p^{2}\right)=\cancel{p}\Sigma_{f}^{V}\left(p^{2}\right)+\cancel{p}\gamma_{5}\Sigma_{f}^{A}\left(p^{2}\right)+M_{f}\Sigma_{f}^{S}\left(p^{2}\right)+M_{f}\gamma_{5}\Sigma_{f}^{P}\left(p^{2}\right). \label{eq:mf-se}
\end{eqnarray} 
The function Dp$\left[\:f\:\right]$ takes the divergent part of the argument $f$. By its own side renormalization of the squark sector requires a more detail description. The necessary counter-terms emerge from the mass term of the bare squark Lagrangian, which is given by
\begin{eqnarray}
\mathcal{L}_{\tilde{m}_{f}}^{\rm{bare}}=-\frac{1}{2}\left(\begin{array}{cc}
\tilde{f}_{L}^{\dagger} & \tilde{f}_{R}^{\dagger}\end{array}\right)M_{L,R}^{2}\left(\begin{array}{c}
\tilde{f}_{L}\\
\tilde{f}_{R}
\end{array}\right), \label{eq:sqL-bare}
\end{eqnarray}
where
\begin{eqnarray}
M_{L,R}^{2}=\left(\begin{array}{cc}
M_{\tilde{U}_L}^{2}+M_{Z}^{2}cos2\beta\left(I_{3}^{f}-Q_{f}s_{W}^{2}\right)+M_{f}^{2} & M_{f}\left(A_{f}-\mu cot\beta\right)\\
M_{f}\left(A_{f}-\mu cot\beta\right) & M_{\tilde{U}_R}^{2}+M_{Z}^{2}cos2\beta Q_{f}s_{W}^{2}+M_{f}^{2}
\end{array}\right) \label{eq:sq-m}
\end{eqnarray}
for the squarks type up. An analogous expression for the squarks type down can be obtained just by changing $cot\beta\rightarrow tan\beta$ and the u-type soft SUSY breaking term $M_{\tilde{U}}^{2}$ by the d-type term $M_{\tilde{D}}^{2}$. The physical mass eigenstates $\tilde{m}_{f_1}$ and $\tilde{m}_{f_2}$  can be found by diagonalizing the mass matrix (\ref{eq:sq-m}) through the orthogonal transformation 
\begin{eqnarray}
\left(\begin{array}{c}
\tilde{f}_{1}\\
\tilde{f}_{2}
\end{array}\right)=U\left(\theta_{\tilde{f}}\right)\left(\begin{array}{c}
\tilde{f}_{L}\\
\tilde{f}_{R}
\end{array}\right) &  ; &  U\left(\theta_{\tilde{f}}\right)=\left(\begin{array}{cc}
c_{\theta_{\tilde{f}}} & s_{\theta_{\tilde{f}}}\\
-s_{\theta_{\tilde{f}}} & c_{\theta_{\tilde{f}}}
\end{array}\right).
\end{eqnarray}
In order to derive the counter-term expressions it is convenient to express the squark mass matrix in terms of the physical masses and the squark mixing angle $\theta_{\tilde{f}}$ as follows:
\begin{eqnarray}
M_{L,R}^{2}=\left(\begin{array}{cc}
c_{\theta_{\tilde{f}}}^{2}\tilde{m}_{f_{1}}^{2}+s{}_{\theta_{\tilde{f}}}^{2}\tilde{m}_{f_{2}}^{2} & s_{\theta_{\tilde{f}}}c_{\theta_{\tilde{f}}}\left(\tilde{m}_{f_{1}}^{2}-\tilde{m}_{f_{2}}^{2}\right)\\
s_{\theta_{\tilde{f}}}c_{\theta_{\tilde{f}}}\left(\tilde{m}_{f_{1}}^{2}-\tilde{m}_{f_{2}}^{2}\right) & c_{\theta_{\tilde{f}}}^{2}\tilde{m}_{f_{2}}^{2}+s{}_{\theta_{\tilde{f}}}^{2}\tilde{m}_{f_{1}}^{2}
\end{array}\right). \label{eq:sqm-diag}
\end{eqnarray}
The renormalization constants of the masses, the mixing angles and the fields  are then defined via the transformations:
\begin{eqnarray}
\tilde{m}_{f_{j}}^{2}\rightarrow\tilde{m}_{f_{j}}^{2}+\delta\tilde{m}_{f_{j}}^{2} &  ;  & \delta\tilde{m}_{f_{j}}^{2}=\sum_{l}\delta^{(l)}\tilde{m}_{f_{j}}^{2},\\ 
\theta_{\tilde{f}}\rightarrow\theta_{\tilde{f}}+\delta\theta_{\tilde{f}} &  ; &  \delta\theta_{\tilde{f}}=\sum_{l}\delta^{(l)}\theta_{\tilde{f}},\\
\left(\begin{array}{c}
\tilde{f}_{1}\\ 
\tilde{f}_{2}
\end{array}\right)\rightarrow Z_{\tilde{f}_{12}}\left(\begin{array}{c}
\tilde{f}_{1}\\
\tilde{f}_{2}
\end{array}\right) &  ; &  Z_{\tilde{f}_{12}}=\left(\begin{array}{cc}
1+\frac{1}{2}\sum_{l}\delta^{(l)}Z_{\tilde{f}_{11}} & \frac{1}{2}\sum_{l}\delta^{(l)}Z_{\tilde{f}_{12}}\\
\frac{1}{2}\sum_{l}\delta^{(l)}Z_{\tilde{f}_{21}} & 1+\frac{1}{2}\sum_{l}\delta^{(l)}Z_{\tilde{f}_{22}}
\end{array}\right),
\end{eqnarray}
where
\begin{eqnarray}
\delta^{(l)}Z_{\tilde{f}_{12}}=\delta^{(l)}Z_{\tilde{f}_{21}}=\frac{s_{\theta_{\tilde{f}}}c_{\theta_{\tilde{f}}}}{c_{\theta_{\tilde{f}}}^{2}-s{}_{\theta_{\tilde{f}}}^{2}}\left(\delta^{(l)}Z_{\tilde{f}_{22}}-\delta^{(l)}Z_{\tilde{f}_{11}}\right) & ; & l=1,2.
\end{eqnarray}
The bare Lagrangian~(\ref{eq:sqL-bare}) is therefore transformed to its renormalized version   
\begin{eqnarray}
\mathcal{L}_{\tilde{m}_{f}}^{{\rm ren}}=-\frac{1}{2}\left(\begin{array}{cc}
\tilde{f}_{1}^{\dagger} & \tilde{f}_{2}^{\dagger}\end{array}\right)Z_{\tilde{f}_{12}}^{T}M_{1,2}^{2}Z_{\tilde{f}_{12}}\left(\begin{array}{c}
\tilde{f}_{1}\\
\tilde{f}_{2}
\end{array}\right)-\frac{1}{2}\left(\begin{array}{cc}
\tilde{f}_{1}^{\dagger} & \tilde{f}_{2}^{\dagger}\end{array}\right)Z_{\tilde{f}_{12}}^{T}\Delta M_{1,2}^{2}Z_{\tilde{f}_{12}}\left(\begin{array}{c}
\tilde{f}_{1}\\
\tilde{f}_{2}
\end{array}\right),\label{eq:sqL-ren}
\end{eqnarray}
with
\begin{eqnarray}
\Delta M_{1,2}^{2}=\left(\begin{array}{cc}
\delta\tilde{m}_{f_{1}}^{2}-\left(\delta\theta_{\tilde{f}}\right)^{2}\left(\tilde{m}_{f_{1}}^{2}-\tilde{m}_{f_{2}}^{2}\right) & \left(\tilde{m}_{f_{1}}^{2}-\tilde{m}_{f_{2}}^{2}\right)\delta\theta_{\tilde{f}}+\delta\theta_{\tilde{f}}\left(\delta\tilde{m}_{f_{1}}^{2}-\delta\tilde{m}_{f_{2}}^{2}\right)\\
\left(\tilde{m}_{f_{1}}^{2}-\tilde{m}_{f_{2}}^{2}\right)\delta\theta_{\tilde{f}}+\delta\theta_{\tilde{f}}\left(\delta\tilde{m}_{f_{1}}^{2}-\delta\tilde{m}_{f_{2}}^{2}\right) & \delta\tilde{m}_{f_{2}}^{2}+\left(\delta\theta_{\tilde{f}}\right)^{2}\left(\tilde{m}_{f_{1}}^{2}-\tilde{m}_{f_{2}}^{2}\right)
\end{array}\right).
\end{eqnarray}
Collecting the counter-terms generated from (\ref{eq:sqL-ren}) and from the kinetic terms of the squark Lagrangian, we derive the expressions of the renormalized squark self-energies, which are given by 
\begin{eqnarray}
\left.\widehat{\sum}\left(p^{2}\right)\right.=Z_{\tilde{f}_{12}}^{\dagger}p^{2}Z_{\tilde{f}_{12}}-Z_{\tilde{f}_{12}}^{\dagger}\left[M_{1,2}^{2}+\Delta M_{1,2}^{2}-\sum\left(p^{2}\right)\right]Z_{\tilde{f}_{12}}-p^{2}+M_{1,2}^{2}. \label{eq:sqse-ren}
\end{eqnarray}
Once again we have used the letter sigma with a hat to represent the renormalized self-energies. Keeping the terms up to two-loop order in the unrenormalized self-energy matrix 
\begin{eqnarray}
\sum\left(p^{2}\right)=\left(\begin{array}{cc}
\sum_{11}^{(1)}+\sum_{11}^{(2)}-2\delta^{(1)}\theta_{\tilde{f}}\left.\sum\right._{12}^{(1)} & \sum_{12}^{(1)}+\sum_{12}^{(2)}+\delta^{(1)}\theta_{\tilde{f}}\left(\left.\sum\right._{11}^{(1)}-\left.\sum\right._{22}^{(1)}\right)\\
\sum_{21}^{(1)}+\sum_{21}^{(2)}+\delta^{(1)}\theta_{\tilde{f}}\left(\left.\sum\right._{11}^{(1)}-\left.\sum\right._{22}^{(1)}\right) & \sum_{22}^{(1)}+\sum_{22}^{(2)}+2\delta^{(1)}\theta_{\tilde{f}}\left.\sum\right._{12}^{(1)}
\end{array}\right)
\end{eqnarray}
as well as in $\Delta M_{1,2}^{2}$ it is possible to derived the components of (\ref{eq:sqse-ren}) at one and two -loop level:
\begin{eqnarray}
\left.\widehat{\sum}\right._{ij}^{(1)}\left(p^{2}\right)=\left.\sum\right._{ij}^{(1)}\left(p^{2}\right)+p^{2}\delta^{(1)}Z_{\tilde{f}_{ij}}-\frac{1}{2}\left(\tilde{m}_{f_{i}}^{2}+\tilde{m}_{f_{j}}^{2}\right)\delta^{(1)}Z_{\tilde{f}_{ij}} \nonumber \\ 
~-~\delta^{(1)}\tilde{m}_{f_{i}}^{2}\delta_{j}^{i}~-~\left(-1\right)^{i+1}\left(\tilde{m}_{f_{i}}^{2}-\tilde{m}_{f_{j}}^{2}\right)\delta^{(1)}\theta_{\tilde{f}} &  ; &  i,j=1,2 \: ,
\end{eqnarray}
\begin{eqnarray}
\left.\widehat{\sum}\right._{ii}^{(2)}\left(p^{2}\right)=\left.\sum\right._{ii}^{(2)}\left(p^{2}\right)+\delta^{(1)}Z_{\tilde{f}_{ii}}\left.\sum\right._{ii}^{(1)}\left(p^{2}\right)+p^{2}\delta^{(2)}Z_{ii}\nonumber\\
-(-1)^{i+1}2\delta^{(1)}\theta_{\tilde{f}}\left.\sum\right._{12}^{(1)}-\tilde{m}_{f_{i}}^{2}\delta^{(2)}Z_{ii}-\delta^{(2)}\tilde{m}_{f_{i}}^{2} \nonumber \\ - \delta^{(1)}\tilde{m}_{f_{i}}^{2}\delta^{(1)}Z_{\tilde{f}_{ii}} - \left(-1\right)^{i}\left(\delta^{(1)}\theta_{\tilde{f}}\right)^{2}\left(\tilde{m}_{f_{1}}^{2}-\tilde{m}_{f_{2}}^{2}\right) &  ; &  i=1,2 \: ,
\end{eqnarray}
\begin{eqnarray}
\left.\widehat{\sum}\right._{12}^{(2)}\left(p^{2}\right)=\left.\sum\right._{12}^{(2)}\left(p^{2}\right)+\frac{1}{2}\left(\delta^{(1)}Z_{\tilde{f}_{11}}+\delta^{(1)}Z_{\tilde{f}_{22}}\right)\left.\sum\right._{12}^{(1)}\left(p^{2}\right) \nonumber \\
~+~\delta^{(1)}\theta_{\tilde{f}}\left(\left.\sum\right._{11}^{(1)}\left(p^{2}\right)-\left.\sum\right._{22}^{(1)}\left(p^{2}\right)\right)-\left(\tilde{m}_{f_{1}}^{2}-\tilde{m}_{f_{2}}^{2}\right)\delta^{(2)}\theta_{\tilde{f}} \nonumber \\
-\left(\tilde{m}_{f_{1}}^{2}-\tilde{m}_{f_{2}}^{2}\right)\delta^{(1)}Z_{\tilde{f}_{11}}\delta^{(1)}\theta_{\tilde{f}} - \left(\delta^{(1)}\tilde{m}_{f_{1}}^{2}-\delta^{(1)}\tilde{m}_{f_{2}}^{2}\right)\delta^{(1)}\theta_{\tilde{f}}.
\end{eqnarray}
As we are working at two-loop order in the $\overline{DR}$ scheme, the left-handed and right-handed components of the squark fields have the same renormalization constants, as a consequence $\delta^{(l)}Z_{\tilde{f}_{11}}=\delta^{(l)}Z_{\tilde{f}_{22}}$ and therefore we have used~$\delta^{(l)}Z_{\tilde{f}_{21}}=0$ in the derivation of the above two-loop expressions. At one-loop level we have imposed the conditions
\begin{eqnarray}
{\rm Dp}\left[\left.\widehat{\sum}\right._{ii}^{(1)}\left(\tilde{m}_{f_{i}}^{2}\right)\right]=0 &  ; &  i=1,2 \: , \\ 
{\rm Dp}\left[\left.\widehat{\sum}\right._{12}^{(1)}\left(\tilde{m}_{f_{2}}^{2}\right)+\left.\widehat{\sum}\right._{21}^{(1)}\left(\tilde{m}_{f_{1}}^{2}\right)\right]=0 &  , & \\ 
{\rm Dp}\left[\frac{\partial}{\partial p^{2}}\left.\widehat{\sum}\right._{ii}^{(1)}\left(\tilde{m}_{f_{i}}^{2}\right)\right]=0 &  ; &  i=1,2 \: ,
\end{eqnarray}
leading to the renormalization constants:
\begin{eqnarray}
\delta^{(1)}\tilde{m}_{f_{i}}^{2}={\rm Dp}\left[\left.\sum\right._{ii}^{(1)}\left(\tilde{m}_{f_{i}}^{2}\right)\right] &  ; &  i=1,2\: ,\\ 
\delta^{(1)}\theta_{\tilde{f}}=\frac{{\rm Dp}\left[\left.\sum\right._{12}^{(1)}\left(\tilde{m}_{f_{2}}^{2}\right)+\left.\sum\right._{21}^{(1)}\left(\tilde{m}_{f_{1}}^{2}\right)\right]}{2\left(\tilde{m}_{f_{1}}^{2}-\tilde{m}_{f_{2}}^{2}\right)} &  , &\\ 
\delta^{(1)}Z_{\tilde{f}_{ii}}=-{\rm Dp}\left[\frac{\partial}{\partial p^{2}}\left.\sum\right._{ii}^{(1)}\left(\tilde{m}_{f_{i}}^{2}\right)\right] &  ; &  i=1,2\: .
\end{eqnarray}
At two-loop order the renormalization conditions are derived by imposing 
\begin{eqnarray}
\left[ p^{2}-M_{1,2}^{2}-\left.\widehat{\sum}\right._{1,2}\right]_{fp}=0,
\end{eqnarray}
that is to say, the finite part of the inverse propagator must be zero. Explicitly the two-loop conditions are:
\begin{eqnarray}
\left[\left.\sum\right._{ii}^{(2)}+\delta^{(1)}Z_{\tilde{f}_{ii}}\left.\sum\right._{ii}^{(1)}+\left(p^{2}-\tilde{m}_{f_{i}}^{2}\right)\delta^{(2)}Z_{ii}-\delta^{(2)}\tilde{m}_{f_{i}}^{2} -\delta^{(1)}\tilde{m}_{f_{i}}^{2}\delta^{(1)}Z_{\tilde{f}_{ii}} \right. \nonumber \\ 
\left.~-~(-1)^{i+1}2\delta^{(1)}\theta_{\tilde{f}}\left.\sum\right._{12}^{(1)}+\left(\delta^{(1)}\theta_{\tilde{f}}\right)^{2}\left(\tilde{m}_{f_{1}}^{2}-\tilde{m}_{f_{2}}^{2}\right)\right]_{fp}=0 \: \: ; \: \: i=1,2\: , \label{eq:2lpoleeqA}
\end{eqnarray} 
\begin{eqnarray}
\left[\left.\sum\right._{12}^{(2)}+\delta^{(1)}Z_{\tilde{f}_{11}}\left.\sum\right._{12}^{(1)}+\delta^{(1)}\theta_{\tilde{f}}\left(\left.\sum\right._{11}^{(1)}-\left.\sum\right._{22}^{(1)}\right)-\left(\tilde{m}_{f_{1}}^{2}-\tilde{m}_{f_{2}}^{2}\right)\delta^{(2)}\theta_{\tilde{f}}\right.\nonumber \\ 
\left.-\left(\delta^{(1)}\tilde{m}_{f_{1}}^{2}-\delta^{(1)}\tilde{m}_{f_{2}}^{2}\right)\delta^{(1)}\theta_{\tilde{f}}-\left(\tilde{m}_{f_{1}}^{2}-\tilde{m}_{f_{2}}^{2}\right)\delta^{(1)}Z_{\tilde{f}_{11}}\delta^{(1)}\theta_{\tilde{f}}\right]_{fp}=0.\label{eq:2lpoleeqB}
\end{eqnarray}
The corresponding two-loop counter-terms in the $\overline{DR}$ scheme are: 
\begin{eqnarray}
\delta^{(2)}\tilde{m}_{f_{j}}^{2}={\rm Dp}\left[\left.\sum\right._{jj}^{(2)}\left(\tilde{m}_{f_{j}}^{2}\right)\right]-(-1)^{j+1}\frac{{\rm Dp}\left[\left.\sum\right._{21}^{(1)}\left(\tilde{m}_{f_{j}}^{2}\right)\right]^{2}}{\left(\tilde{m}_{f_{1}}^{2}-\tilde{m}_{f_{2}}^{2}\right)} \nonumber \\ 
~+~~(-1)^{j+1}\frac{{\rm Dp}\left[\left.\sum\right._{12}^{(1)}\left(\tilde{m}_{f_{2}}^{2}\right)-\left.\sum\right._{21}^{(1)}\left(\tilde{m}_{f_{1}}^{2}\right)\right]^{2}}{4\left(\tilde{m}_{f_{1}}^{2}-\tilde{m}_{f_{2}}^{2}\right)} &  ; &  j=1,2\: , \label{eq:ct-d2msf}
\end{eqnarray}
\begin{eqnarray}
2\left(\tilde{m}_{f_{1}}^{2}-\tilde{m}_{f_{2}}^{2}\right)\delta^{(2)}\theta_{\tilde{f}}~=~~{\rm Dp}\left[\left.\sum\right._{12}^{(2)}\left(\tilde{m}_{f_{1}}^{2}\right)\right]+{\rm Dp}\left[\left.\sum\right._{12}^{(2)}\left(\tilde{m}_{f_{2}}^{2}\right)\right] \nonumber \\ 
-~{\rm Dp}\left[\delta^{(1)}\theta_{\tilde{f}}\right]{\rm Dp}\left[\left.\sum\right._{11}^{(1)}\left(\tilde{m}_{f_{2}}^{2}\right)-\left.\sum\right._{11}^{(1)}\left(\tilde{m}_{f_{1}}^{2}\right)\right] \nonumber \\ 
-~{\rm Dp}\left[\delta^{(1)}\theta_{\tilde{f}}\right]{\rm Dp}\left[\left.\sum\right._{22}^{(1)}\left(\tilde{m}_{f_{1}}^{2}\right)-\left.\sum\right._{22}^{(1)}\left(\tilde{m}_{f_{2}}^{2}\right)\right] &  , & \label{eq:ct-d2tht}
\end{eqnarray}
\begin{eqnarray}
\delta^{(2)}Z_{ii}=-~~{\rm Dp}\left[\frac{\partial}{\partial p^{2}}\left.\sum\right._{ii}^{(2)}\left(\tilde{m}_{f_{i}}^{2}\right)\right]+{\rm Dp}\left[\frac{\partial}{\partial p^{2}}\left.\sum\right._{ii}^{(1)}\left(\tilde{m}_{f_{i}}^{2}\right)\right]^{2} \nonumber \\ 
+~~(-1)^{i+1}2{\rm Dp}\left[ \delta^{(1)}\theta_{\tilde{f}} \right] {\rm Dp}\left[\frac{\partial}{\partial p^{2}}\left.\sum\right._{12}^{(1)}\left(\tilde{m}_{f_{i}}^{2}\right)\right] &  ; &  i=1,2\: , \label{eq:ct-d2Z}
\end{eqnarray}
In the NLF limit the mixing angles $\theta_{\tilde{q}}$ with $q=u,d,b,c,s$ are equal to zero and therefore there are no mixing counter-terms. Thus, the pole equation (\ref{eq:2lpoleeqB}) and the counter-term (\ref{eq:ct-d2tht}) don't exist for the $q$-type squarks. Additionally, the equation (\ref{eq:2lpoleeqA}) and the counter-terms (\ref{eq:ct-d2msf}) and (\ref{eq:ct-d2Z}) can be even further reduced just by putting
$\delta^{(1)}\theta_{\tilde{q}}=0$.   

\section{\large Renormalization Constants}
\label{ap:ren-const}
In this appendix we are going to list the relevant counter-terms involved in our calculation. The one-loop counter-term of the top quark mass in the $\overline{DR}$ scheme is:
\begin{eqnarray}
\delta^{(1)} M_{t}=\frac{\alpha_{s}}{6\pi}{\rm Dp}\left[A0\left(\tilde{m}_{t_{1}}^{2}\right)+A0\left(\tilde{m}_{t_{2}}^{2}\right)-2A0\left(M_{\tilde{g}}^{2}\right)-2A0\left(M_{t}^{2}\right)-4M_{t}^{2}B0\left(M_{t}^{2},M_{t}^{2},0\right)\right. \nonumber \\ 
\left. ~+~ \left(M_{\tilde{g}}^{2}+M_{t}^{2}-\tilde{m}_{t_{1}}^{2}\right)B0\left(M_{t}^{2},\tilde{m}_{t_{1}}^{2},M_{\tilde{g}}^{2}\right)+\left(M_{\tilde{g}}^{2}+M_{t}^{2}-\tilde{m}_{t_{2}}^{2}\right)B0\left(M_{t}^{2},\tilde{m}_{t_{2}}^{2},M_{\tilde{g}}^{2}\right)\right. \nonumber \\ 
\left. ~+~~ 2M_{\tilde{g}}M_{t}s_{2\theta_{\tilde{t}}}\left(B0\left(M_{t}^{2},\tilde{m}_{t_{2}}^{2},M_{\tilde{g}}^{2}\right)-B0\left(M_{t}^{2},\tilde{m}_{t_{1}}^{2},M_{\tilde{g}}^{2}\right)\right)\right] \: . 
\end{eqnarray}
The gluino mass counter-term at one-loop order is:
\begin{eqnarray}
\delta^{(1)}M_{\tilde{g}}=\frac{\alpha_{s}}{8\pi}{\rm Dp}\left[\sum_{f}\sum_{j=1}^{2}A0\left(\tilde{m}_{f_{j}}^{2}\right)-6A0\left(M_{\tilde{g}}^{2}\right)-2A0\left(M_{t}^{2}\right)-12M_{\tilde{g}}^{2}B0\left(M_{\tilde{g}}^{2},M_{\tilde{g}}^{2},0\right)\right. + \nonumber \\ 
\left. \sum_{f}\sum_{j=1}^{2}\left(M_{\tilde{g}}^{2}+M_{f}^{2}-\tilde{m}_{f_{j}}^{2}-2M_{\tilde{g}}M_{f}s_{2\theta_{\tilde{f}}}\right)B0\left(M_{\tilde{g}}^{2},\tilde{m}_{f_{j}}^{2},M_{f}^{2}\right)\right] \:\: ; \: f=u,d,t,b,c,s \: .
\end{eqnarray}
In the NLF limit $M_q=0$ where $q=u,d,b,c,s$. From now on we are going to preserve the definitions of the indices $f$ and $q$. \\ The counter-terms of the squark sector are:
\begin{eqnarray}
\delta^{(1)}\tilde{m}_{f_{1}}^{2}=\frac{\alpha_{s}}{3\pi}{\rm Dp}\left[A0\left(\tilde{m}_{f_{1}}^{2}\right)c_{2\theta_{\tilde{f}}}^{2}+A0\left(\tilde{m}_{f_{2}}^{2}\right)s_{2\theta_{\tilde{f}}}^{2}+A0\left(\tilde{m}_{f_{1}}^{2}\right)-4\tilde{m}_{f_{1}}^{2}B0\left(\tilde{m}_{f_{1}}^{2},\tilde{m}_{f_{1}}^{2},0\right)\right]~-~ \nonumber \\ 
2\frac{\alpha_{s}}{3\pi}{\rm Dp}\left[A0\left(M_{\tilde{g}}^{2}\right)+A0\left(M_{f}^{2}\right)+\left(M_{\tilde{g}}^{2}+M_{f}^{2}-\tilde{m}_{f_{1}}^{2}+(-1)^{1}2M_{\tilde{g}}M_{f}s_{2\theta_{\tilde{f}}}\right)B0\left(\tilde{m}_{f_{1}}^{2},M_{f}^{2},M_{\tilde{g}}^{2}\right)\right] \: .
\end{eqnarray}
The expression for $\delta^{(1)}\tilde{m}_{f_{2}}^{2}$ is obtained by interchanging $1\leftrightarrow 2$ in the above equation. Besides
\begin{eqnarray}
\delta^{(1)}\theta_{\tilde{t}}=\left\{ 4\frac{\alpha_{s}}{3\pi}M_{\tilde{g}}M_{t}\left(c_{2\theta_{\tilde{t}}}^{2}-s_{2\theta_{\tilde{t}}}^{2}\right){\rm Dp}\left[B0\left(\tilde{m}_{t_{1}}^{2},M_{t}^{2},M_{\tilde{g}}^{2}\right)+B0\left(\tilde{m}_{t_{2}}^{2},M_{t}^{2},M_{\tilde{g}}^{2}\right)\right]\right. \nonumber \\ 
\left.-\frac{\alpha_{s}}{3\pi}{\rm Dp}\left[A0\left(\tilde{m}_{t_{1}}^{2}\right)-A0\left(\tilde{m}_{t_{2}}^{2}\right)\right]s_{4\theta_{\tilde{t}}}\right\} \left/2\left(\tilde{m}_{t_{1}}^{2}-\tilde{m}_{t_{2}}^{2}\right)\right. \: , 
\end{eqnarray}
and
\begin{eqnarray}
\delta^{(1)}Z_{\tilde{f}_{jj}}=\frac{\alpha_{s}}{3\pi}{\rm Dp}\left[2\left(M_{\tilde{g}}^{2}+M_{f}^{2}-\tilde{m}_{f_{j}}^{2}+(-1)^{j}2M_{\tilde{g}}M_{f}s_{2\theta_{\tilde{f}}}\right)\frac{\partial}{\partial p^{2}}B0\left(p^{2},M_{f}^{2},M_{\tilde{g}}^{2}\right)\right. \nonumber \\ 
\left.\left.-2B0\left(\tilde{m}_{f_{j}}^{2},M_{f}^{2},M_{\tilde{g}}^{2}\right)+4\tilde{m}_{f_{j}}^{2}\frac{\partial}{\partial p^{2}}B0\left(p^{2},\tilde{m}_{f_{j}}^{2},0\right)+2B0\left(\tilde{m}_{f_{j}}^{2},\tilde{m}_{f_{j}}^{2},0\right)\right]\right|_{p^{2}=\tilde{m}_{f_{j}}^{2}}; \:\: j=1,2 \: .
\end{eqnarray}
The one-loop Passarino-Veltman functions $A0$ and $B0$ have the Laurent expansions:
\begin{eqnarray}
 A0\left(m^{2}\right) = \frac{e^{\gamma_{E}\varepsilon}}{i\pi^{D/2}}\int d^{D}q\frac{1}{q^{2}-m^{2}} \;\; \qquad \nonumber \\  = -e^{\gamma_{E}\varepsilon}\left(m^{2}\right)^{1-\varepsilon}\Gamma\left(-1+\varepsilon\right) \: ,
\end{eqnarray}
and
\begin{eqnarray}
B0\left(p^{2},m_{1}^{2},m_{2}^{2}\right)= \frac{e^{\gamma_{E}\varepsilon}}{i\pi^{D/2}}\int d^{D}q\frac{1}{\left[q^{2}-m_{1}^{2}\right]\left[\left(q+p\right)^{2}-m_{2}^{2}\right]} \qquad\qquad\qquad\qquad\qquad\qquad\qquad \nonumber \\  = \left(p^{2}\right)^{-\varepsilon}\left[\frac{1}{\varepsilon}+2-Log\left(\frac{m_{1}m_{2}}{p^{2}}\right)+\frac{m_{1}^{2}-m_{2}^{2}}{p^{2}}Log\left(\frac{m_{2}^{2}}{m_{1}^{2}}\right)+\lambda\left(1,\frac{m_{1}^{2}}{p^{2}},\frac{m_{2}^{2}}{p^{2}}\right)\times\right. \nonumber \\ 
\left(\left. Log\left(\frac{m_{1}m_{2}}{p^{2}}\right)-Log\left(\frac{p^{2}-m_{1}^{2}-m_{2}^{2}+\lambda\left(1,m_{1}^{2}/p^{2},m_{2}^{2}/p^{2}\right)}{2p^{2}}\right)+i\pi\right)\right]+O\left(\varepsilon\right), 
\end{eqnarray}
where $\lambda(x,y,z)$ is the K\"{a}llen function. For a complex number $z$ the logarithm is defined as $Log(z) = ln|z| + i\phi$, where $\phi~\in~(-\pi,\pi]$. \\ 
At two-loop level we need the $\overline{DR}$ counter-terms:
\begin{eqnarray}
\small
\delta^{(2)}M_{t}=\left(\frac{\alpha_{s}}{4\pi}\right)^{2}M_{t}^{2}\left[\frac{68}{9}\frac{1}{\varepsilon^{2}}-\frac{4}{9}\frac{1}{\varepsilon}\right], \label{eq:d2Mt}
\end{eqnarray}
\begin{eqnarray}
\delta^{(2)}\tilde{m}_{t_{1}}^{2}=\left(\frac{\alpha_{s}}{\pi}\right)^{2}\left\{ \frac{16}{9}c_{2\theta_{\tilde{t}}}^{2}\frac{M_{\tilde{g}}^{2}M_{t}^{2}}{\left(\tilde{m}_{t_{1}}^{2}-\tilde{m}_{t_{2}}^{2}\right)}+\left(\frac{1}{9}\left(1+c_{2\theta_{\tilde{t}}}^{2}\right)+\frac{1}{8}\right)s_{2\theta_{\tilde{t}}}^{2}\left(\tilde{m}_{t_{1}}^{2}-\tilde{m}_{t_{2}}^{2}\right)\right. \nonumber \\ 
\left.~+~~\frac{25}{18}M_{t}^{2}+\frac{3}{2}M_{\tilde{g}}^{2}-\left(\frac{8}{9}\left(1+c_{2\theta_{\tilde{t}}}^{2}\right)+1\right)s_{2\theta_{\tilde{t}}}M_{\tilde{g}}M_{t}\right\} \frac{1}{\varepsilon^{2}} \nonumber \\ 
~+~~\left(\frac{\alpha_{s}}{\pi}\right)^{2}\left\{ \frac{1}{6}\left(\tilde{m}_{t_{1}}^{2}+\tilde{m}_{t_{2}}^{2}\right)-\frac{1}{72}s_{2\theta_{\tilde{t}}}^{2}\left(\tilde{m}_{t_{1}}^{2}-\tilde{m}_{t_{2}}^{2}\right)-\frac{7}{18}M_{t}^{2}\right. \nonumber \\ 
\left. ~-~\frac{7}{6}M_{\tilde{g}}^{2}~+~\frac{1}{9}s_{2\theta_{\tilde{t}}}M_{\tilde{g}}M_{t}~+~\frac{1}{6}\sum_{q}\sum_{j=1}^{2}\tilde{m}_{q_{j}}^{2}-\frac{1}{2}m_{\varepsilon}^{2}\right\} \frac{1}{\varepsilon} \: .  \label{eq:d2mt1exp}
\end{eqnarray}
The expression for $\delta^{(2)}\tilde{m}_{t_2}$ can be derived from (\ref{eq:d2mt1exp}) by interchanging the indices $1 \leftrightarrow 2$ and changing the sign of $\theta_{\tilde{t}}$. Furthermore in the $\overline{DR}$ scheme the two-loop counter-terms of the stop masses have a dependence on the $\varepsilon$-scalar mass $m_{\varepsilon}$. This dependence cancels in the final result of our three-loop correction to $M_h$ if we renormalize the $\varepsilon$-scalar mass in the  $\overline{DR}$ scheme. It is enough to consider the one-loop counter-term
\begin{eqnarray}
\delta^{(1)}m_{\varepsilon}^{2}=\frac{\alpha_{s}}{\pi}m_{\varepsilon}^{2}\left\{ -\frac{3}{4}+\left[-\frac{3}{2}M_{\tilde{g}}^{2}+\frac{1}{2}\sum_{f}\sum_{j=1}^{2}\tilde{m}_{f_{j}}^{2}-M_{t}^{2}\right]\frac{1}{2m_{\varepsilon}^{2}}\right\} \frac{1}{\varepsilon} \: .
\end{eqnarray}
Finally, we need the two-loop counter-term of the stop mixing angle:  
\begin{eqnarray}
\left(\tilde{m}_{t_{1}}^{2}-\tilde{m}_{t_{2}}^{2}\right)\delta^{(2)}\theta_{\tilde{t}}=\left(\frac{\alpha_{s}}{\pi}\right)^{2}\left\{ \left(\frac{1}{8}s_{2\theta_{\tilde{t}}}c_{2\theta_{\tilde{t}}}-\frac{1}{9}\left(s_{2\theta_{\tilde{t}}}^{2}-c_{2\theta_{\tilde{t}}}^{2}\right)\right)s_{2\theta_{\tilde{t}}}\left(\tilde{m}_{t_{1}}^{2}-\tilde{m}_{t_{2}}^{2}\right)\right.~ + ~ \nonumber \\ 
\left.\left(\frac{8}{9}c_{2\theta_{\tilde{t}}}\left(s_{2\theta_{\tilde{t}}}^{2}-c_{2\theta_{\tilde{t}}}^{2}\right)-c_{2\theta_{\tilde{t}}}\right)M_{\tilde{g}}M_{t}-\frac{32}{9}s_{2\theta_{\tilde{t}}}c_{2\theta_{\tilde{t}}}\frac{M_{\tilde{g}}^{2}M_{t}^{2}}{\left(\tilde{m}_{t_{1}}^{2}-\tilde{m}_{t_{2}}^{2}\right)}\right\} \frac{1}{\varepsilon^{2}} \nonumber \\ 
+\:\:\left(\frac{\alpha_{s}}{\pi}\right)^{2}\left\{ \frac{1}{9}c_{2\theta_{\tilde{t}}}M_{\tilde{g}}M_{t}-\frac{1}{72}s_{2\theta_{\tilde{t}}}c_{2\theta_{\tilde{t}}}\left(\tilde{m}_{t_{1}}^{2}-\tilde{m}_{t_{2}}^{2}\right)\right\} \frac{1}{\varepsilon}\: .
\end{eqnarray}

\end{document}